\documentclass{IEEEtran}
\usepackage{amssymb}
\usepackage{amsmath}
\usepackage{graphicx}
\usepackage{enumerate}
\newtheorem{my_theorem}{Theorem}

\newtheorem{my_lemma}{Lemma}

\usepackage{subfigure}
\usepackage{mathtools}
\usepackage{cite}
\usepackage{color}
\usepackage{caption}
\usepackage{color}

\title{Multihop  Optical Wireless Communication Over ${\cal{F}}$-Turbulence Channels and Generalized Pointing Errors with Fog-Induced Fading}
\author{
}

\author{Ziyaur Rahman,~\IEEEmembership{Student Member,~IEEE}, S.~M.~ Zafaruddin,~\IEEEmembership{Senior Member,~IEEE,} and
	V.~K.~ Chaubey,~\IEEEmembership{Senior Member,~IEEE}
	
	\thanks{Ziyaur Rahman (p20170416@pilani.bits-pilani.ac.in), S. M. Zafaruddin (syed.zafaruddin@pilani.bits-pilani.ac.in), and V. K. Chaubey (vkc@pilani.bits-pilani.ac.in)  are with the Department of Electrical and Electronics Engineering, BITS Pilani, Pilani-333031, Rajasthan, India.}

\thanks{This work was supported in part by the 
	Science and Engineering Research Board (SERB), Department of Science and Technology (DST), Govt. of India under MATRICS Grant MTR/2021/000890 and Start-up Research Grant SRG/2019/002345.}}

\begin{document}
\maketitle

\begin{abstract}
Multihop relaying is a potential technique to mitigate channel impairments in   optical wireless communications (OWC).   In this paper,  multiple fixed-gain amplify-and-forward (AF) relays are employed to enhance the OWC performance under the combined effect of atmospheric turbulence, pointing errors, and fog. We consider a long-range OWC link by  modeling the atmospheric turbulence by the Fisher-Snedecor ${\cal{F}}$ distribution,  pointing errors by the generalized non-zero boresight model, and random path loss due to fog. We also consider a short-range OWC system by ignoring  the impact of atmospheric turbulence. We derive novel upper bounds on the probability density function (PDF) and cumulative distribution function (CDF) of the end-to-end signal-to-noise ratio (SNR)  for both  short and long-range multihop OWC systems   by developing  exact statistical results  for a single-hop OWC system under the combined effect of  ${\cal{F}}$-turbulence channels, non-zero boresight pointing errors, and fog-induced fading. Based on these expressions, we present analytical expressions of  outage probability (OP) and average bit-error-rate (ABER) performance for the considered OWC systems  involving single-variate Fox's H and Meijer's G functions.  Moreover,  asymptotic expressions of the outage probability in high SNR region are developed using simpler Gamma functions to provide insights on the effect of channel and system parameters.  The  derived analytical expressions are validated through  Monte-Carlo simulations, and the scaling of the OWC performance with the number of relay nodes is demonstrated with a comparison to  the single-hop transmission.	
\end{abstract}


  \begin{table*}[tp]
  	
  	\caption{Related literature on AF-assisted multihop OWC systems}
  	\label{table:mh_literature}
  	\centering
  	\begin{tabular}{  c c c c c}
  		\hline
  		\hline
  		Reference   & Detector & Turbulence  & Pointing Errors & Fog-Induced Fading   \\
  		\hline
  		[9] &   IM/DD &Gamma-Gamma  &No & No \\  \hline
  	    [18] &   IM/DD & Gamma-Gamma  &No & No \\  \hline
  		[19] &  IM/DD& Gamma-Gamma  &Zero boresight& No \\  \hline
  		[20] &   HD& Gamma-Gamma  &Zero boresight & No \\  \hline
  		[21] &   HD & Gamma-Gamma &Zero boresight & No \\  \hline
  		[22] &   IM/DD & Mal\'aga  &Zero boresight & No \\  \hline
  		[23] &   HD & Mal\'aga &Zero boresight & No \\  \hline
  		[26] & IM/DD& Double Generalized
  		Gamma  &Zero boresight& No \\  \hline
  	  		[This paper] & IM/DD &Fisher-Snedecor ${\cal{F}}$ & Non-zero boresight & Yes\\
  		\hline
  		\hline
  		
  	\end{tabular}
  	
  \end{table*}

  	\begin{table*}[tp]	
	\caption{Related literature on OWC systems with Fog-induced fading}
	\label{table:fog_literature}
  		\centering
  		\begin{tabular}{ c c c c}
  			\hline
  			\hline
  			Reference   & System Model &Pointing Errors & Turbulence\\
  			\hline
  			[34] &  Single-hop& No  &No  \\  \hline
  			[35]&   Single-hop &No &  No \\ \hline
  			[46] &  Single-hop& No & No \\ \hline
  		
  			[37] &  Single-hop & Zero boresight & No  \\ \hline
  				[42]& Single-hop &Zero boresight &Fisher-Snedecor ${\cal{F}}$   \\  \hline
  			[38]&  Dual-hop, DF &Zero boresight &Double Generalized
  			Gamma\\\hline
  		  				[36] &  Multihop, DF & Zero boresight & No  \\ \hline
  			[This paper] & Single and Multihop, AF & Non-zero boresight & Fisher-Snedecor ${\cal{F}}$\\
  			\hline		
  		\end{tabular}	
  	\end{table*}

\section{Introduction}
Optical wireless communication (OWC)  is emerging as a key technology for backhaul connectivity in the next-generation wireless network \cite{Khalighi2014, Kaushal2017, Mohamed2018, Xu2019}. The OWC system exploits  large unlicensed bandwidth to provide  exceedingly higher data rate with low latency transmissions and possesses narrow beam divergence for secured data links without electromagnetic interference. Despite these advantages, the OWC technology is susceptible to atmospheric conditions such as turbulence environment and foggy weather conditions and requires line-of-sight prorogation  with near-perfect beam-alignment between the transmitter and detector.  Developing efficient  techniques to mitigate these channel fading impairments efficiently is desirable for an effective design of OWC systems.

Multi-aperture and cooperative relaying are potential techniques to mitigate the fading effect and extend the communication range. There has been extensive research on  relay-assisted OWC systems with regenerative and non-regenerative protocols under the combined effect of  atmospheric turbulence and pointing errors. Dual-hop relaying using both decode-and-forward (DF) and amplify-and-forward (AF) protocols is a well-studied topic for optical wireless channels \cite{Aggarwal2014,Yang2014,Libich2015,Zedini2017}.  However,  analyzing the performance of  multihop relaying is challenging for mathematically complicated turbulence fading models, especially when the  AF relaying  is employed at each hop. It is known that analyzing DF-assisted multihop system is greatly simplified since the performance analysis decouples into each hop independently and thus  requires   statistical derivation of a single link  \cite{Tsiftsis2006, Aghajanzadeh2011, Safari2012, Kashani2013, Wang2015, Wang2015_ew_pe, Wang2016_NZ,Ben_Issaid2017_NZ}. Further, the DF-based multihop requires channel state information (CSI) at each hop and  becomes impractical when the number of hops increases  beyond a certain limit.  

Scanning the literature, there have been studies on the  AF relaying for  multihop OWC transmissions  \cite{ Tsiftsis2006, Datsikas2010,   Zedini2015, Zedini2015_IEEE_PJ, Tang2014,  Alheadary2017_VTC,Alheadary2017_IWCMC, Pang2021, Ashrafzadeh2018, Ashrafzadeh2020}. Tsiftsis et al.  analyzed the exact performance of multihop optical transmission  over Gamma-Gamma atmospheric turbulence using channel-assisted AF relaying \cite{ Tsiftsis2006}. However,  fixed-gain  relaying is desirable due to a simpler implementation but analyzing its performance becomes intractable for many fading channels. To circumvent this,  the end-to-end  signal-to-noise ratio (SNR) of the  fixed-gain multihop AF relaying is upper bounded as the product of SNR of individual links for tractable performance analysis. This approach was first considered in  \cite{Karagiannidis2006} to analyze the multihop relayed communication over Nakagami-m fading channels. Datsikas et al. analyzed the outage probability (OP) and average bit-error-rate (ABER) of an AF-assisted multihop system over Gamma-Gamma atmospheric turbulence without considering the impact of pointing errors \cite{Datsikas2010}. Tang et al.  extended the work presented in \cite{Datsikas2010}  considering the combined effect of pointing errors and Gamma-Gamma distributed atmospheric turbulence for the heterodyne OWC system \cite{Tang2014}.  Zedini and Alouini  developed exact and asymptotic analysis for the OP,  ABER, and ergodic capacity for an OWC system, pointing errors and atmospheric turbulence.   Later, they  extended the analysis presented in  \cite{ Zedini2015_IEEE_PJ}  for the intensity modulation/decision-directed (IM/DD) OWC system  \cite{Zedini2015}.    Alheadary et al. considered the misaligned generalized Malaga turbulence and developed ABER performance for the multihop heterodyne OWC system \cite{Alheadary2017_IWCMC}.   Ashrafzadeh et al. used the method of induction to develope and  exact analysis  for an AF-assisted multihop OWC system  under the double generalized Gamma turbulence with pointing errors \cite{Ashrafzadeh2018, Ashrafzadeh2020}.

In the above and related literature, we find a few research gaps in the study of the AF-assisted multihop OWC system, which need to be addressed. Firstly,  the AF-assisted multihop transmission was limited to zero-boresight pointing errors and random jitter. The jitter is a random offset of the beam center at the detector plane, and the boresight is the fixed displacement between the center of the beam and the center of the detector \cite{Yang2014}.  The nonzero boresight model  generalizes the effect of pointing errors on the OWC. Although the statistical analysis of  nonzero boresight pointing errors for OWC system has been investigated extensively in the literature for single-link, dual-hop, and multi-aperture, it has not been addressed for the AF-assisted multihop transmission. Indeed, there are a few works on the impact of generalized pointing errors but for DF-assisted multihop systems \cite{Wang2016_NZ,Ben_Issaid2017_NZ}.  

Secondly, adopted models for atmospheric turbulence were    Gamma-Gamma, double generalized Gamma, and Mal\'aga distributions.  These models are mathematically complicated by including Bessel  and Meijer's G-functions in their probability density functions (PDF).  Recently, Peppas et al. proposed a new  atmospheric  turbulence model for OWC, called Fisher-Snedecor $\cal{F}$,  which provides a better fit for experimental data for all turbulence conditions \cite{Peppas2020_F}. Further, the PDF of  $\cal{F}$ turbulence is more mathematically  tractable  since it includes  elementary functions. There has been an increased interest to analyze the performance of OWC over  $\cal{F}$ turbulence
\cite{Badarneh2021_F,Badarneh2020_F2, Han2021_secrecy, Shakir2022_secrecy,Ding2022_F}. However, no work has been reported even with the DF-assisted multihop OWC system considering the $\cal{F}$ atmospheric turbulence model.

Thirdly, signal attenuation for OWC transmissions is assumed to be deterministic and quantified using a visibility range, for example, less attenuation in haze and more loss of signal power in foggy conditions. However, recent measurement data confirm that the signal attenuation in  foggy weather for terrestrial applications is not deterministic but follows a probabilistic model \cite{Khan2009, Esmail2016_Photonics,Esmail2017_Access}. The authors in \cite{Esmail2017_Photonics, rahman2020cl} analyzed the single-link  performance of the OWC system under the combined effect of fog and pointing errors. Esmail et al.   analyzed the OP of a multihop relay system employing  the DF protocol (using the cumulative distribution function (CDF) of the SNR for a single link) to mitigate the effect of fog and pointing errors \cite{Esmail2017_Photonics}. In our previous work \cite{Rahman2021TVT}, we studied the DF-based dual-hop relaying for the OWC system  under the combined effect of random fog,  zero-boresight pointing errors, and atmospheric turbulence distributed according to the double generalized gamma.     To the best of the authors' knowledge,  there are no  analyses available for AF-assisted  multihop OWC system under the effect of random fog, non-zero boresight pointing errors, and atmospheric turbulence. In Table 	\ref{table:mh_literature} and Table \ref{table:fog_literature}, we provide a summary of the state-of-the art research  on the multihop OWC system and OWC system with fog-induced fading, respectively.

In this paper,   we employ multiple fixed-gain relays in each hop to enhance the OWC performance under the combined effect of atmospheric turbulence, pointing errors, and fog. The major contributions of the proposed work are listed as follows:
\begin{itemize}
	\item  We analyze  the end-to-end performance of a fixed-gain AF  multihop relayed OWC system considering independent and non-identical (i.ni.d) fading channels in each hop distributed according to the combined statistics of ${\cal{F}}$-turbulence channels, non-zero boresight pointing errors, and  fog-induced fading.   
	\item We develop exact statistical results  for the direct link of OWC system under the combined effect of atmospheric turbulence, pointing errors, and random fog. We also develop statistical results for  a short-range OWC system by ignoring  the impact of atmospheric turbulence.
	\item  We derive novel upper bounds on the PDF and CDF of the end-to-end SNR for  fixed gain assisted  multihop relaying  for both  short and long-range OWC systems involving single-variate Fox's H and Meijer's G functions.

	\item  We use the derived statistical results to analyze the  OP and ABER performance for single-hop  and multihop transmissions considering both   short and long-range OWC systems. 
	\item We develop  diversity order of the considered system by deriving asymptotic expressions of the OP in high SNR 	region  to provide insights on the design aspects of channel and system parameters.
		\item We use computer simulations to demonstrate the significance of multihop relaying for OWC systems to mitigate the channel impairment compared with the single-hop system.
\end{itemize}

\emph{Notations:} $(\cdot)_{i}$ denotes the $i$-th hop and $(\cdot)_{N}$ denotes the $N$-hop system. We denote the expectation operator by $\mathbb{E}[\cdot]$, Gamma function by $\Gamma(a)=\int\limits_{0}^{\infty}t^{a-1} e^{-t}dt$, upper incomplete Gamma function by  $\Gamma(a,t)=\int_{t}^{\infty}s^{a-1}e^{-s}ds$, Gaussian Q function by  $Q(\gamma)=\frac{1}{\sqrt{2 \pi}}\int_{\gamma}^{\infty}e^{-\frac{u^2}{2}}du$, Meijer's G-function by $G_{p,q}^{m,n}\left[z \middle \vert \begin{matrix}(a_k)_{k=1:p}  \\ (b_k)_{k=1:q}\end{matrix} \right]$, and  the Fox's H-function by 	 $H_{p,q}^{m,n} \left[z \middle \vert \begin{matrix}(a_k,A_k)_{k=1:p}  \\ (b_k,B_k)_{k=1:q}\end{matrix} \right]$.

  The rest of this paper is organized as follows. Section II describes the channel models for fog, non-zero boresight pointing errors, and atmospheric turbulence for multihop OWC communication. In Sections III and IV, the performance of long and short-range OWC systems in terms of OP and ABER is presented, respectively. Numerical and simulation results are presented in Section V.  Finally, we conclude the findings of this paper in Section VI.

\section{System Model}\label{sec:system_model}

\begin{figure}[t]	
	\centering
	\includegraphics[width=\columnwidth]{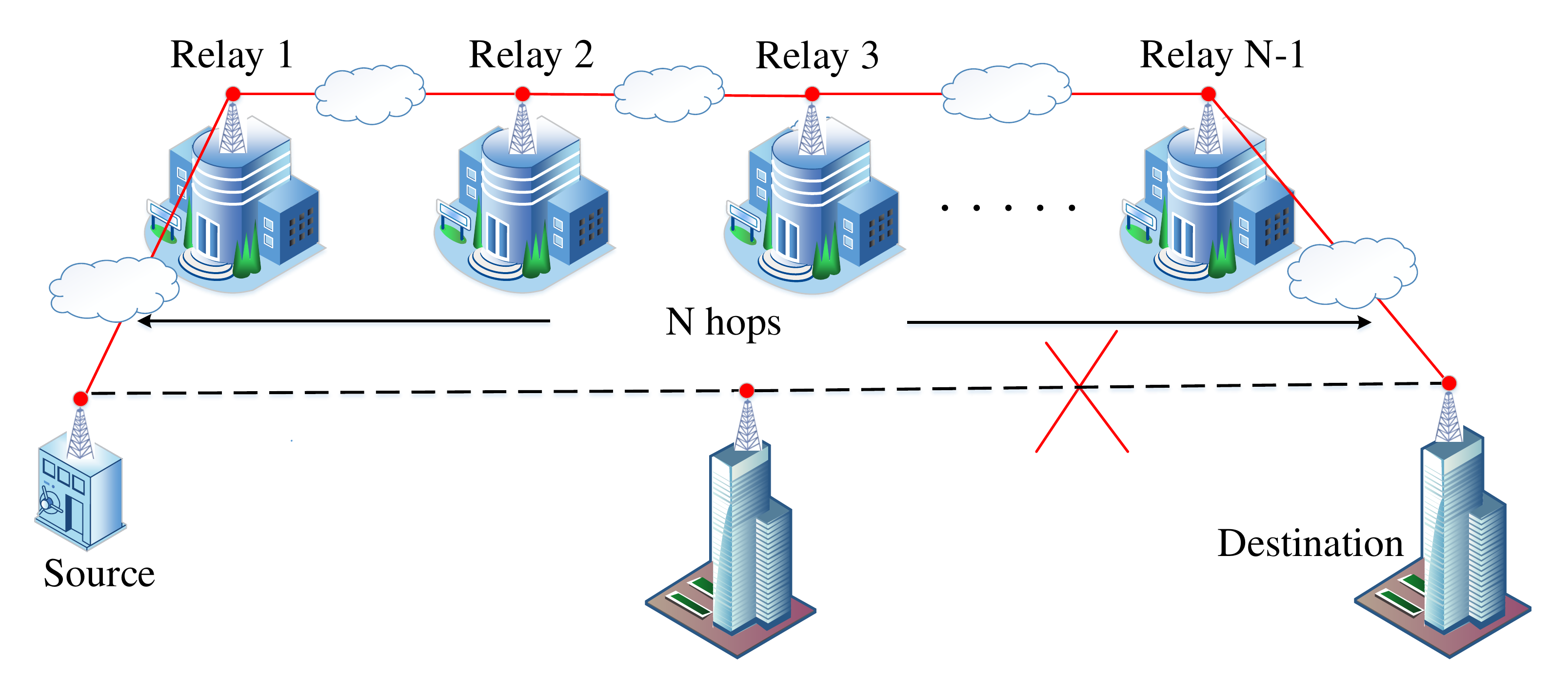}
	\caption{Multihop OWC communication system.}
	\label{system_model}
\end{figure}

 We consider an $N$-hop OWC system where  a source terminal communicates with the destination through $N-1$ relay nodes.   The transmitted signal is impaired by the multiplicative channel effects of  atmospheric turbulence, pointing error, and random fog.  Thus, the received signal $y_i$  for the $i$-th hop  under the additive noise $w_i$ with variance $\sigma^2_{w_i}$ is given by
\begin{flalign}
y_i = h_is + w_i,
\label{eq:received_signal}
\end{flalign}
 where $s$  is the transmitted signal and $h_i=h_{f_i}h_{p_i}h_{t_i}$ is the channel coefficient of the $i$-th hop. Here,   $h_{f_i}$ denotes the random fog,  $h_{p_i}$ denotes the pointing error,  $h_{t_i}$ denote the atmospheric turbulence. We use the simpler non-coherent intensity modulation/direct detection (IM/DD) scheme since  heterodyne detection (HD) requires complex processing of  mixing  the received signal with a coherent signal produced by the local oscillator. Assuming IM/DD technique and on-off keying (OOK) modulation with $x\in \{0,\sqrt{2}P_i\}$  and $P_i$ as the average transmitted optical power, the instantaneous  received electrical SNR for the $i$-th hop is given by \cite{Farid2007}
  \begin{flalign}
  \gamma_i^{\rm SH} = \frac{2P^2_i h_i^2}{\sigma^2_{w_i}} = \bar{\gamma}_ih_i^2
  \label{eq:inst_snr}
  \end{flalign}
where $\bar{\gamma}_i$ is the average SNR of the $i$-th hop. Employing AF relaying with fixed-gain in each hop, the SNR of the $N$-th hop system is given by \cite{Hasna2003}: 
\begin{eqnarray} \label{eq:snr_cg_exact}
	\gamma_{\gamma_N}^{\rm MH} = \Bigg(\sum_{i=1}^{N} \prod_{j=1}^{i} \frac{C_j-1}{\gamma_{j}}\Bigg)^{-1}
\end{eqnarray}
where $C_j$ is constant depending on the AF gain of the $j$-th relay.
Since the SNR expression in \eqref{eq:snr_cg_exact} is intractable for statistical analysis, we use a popular upper bound on $\gamma_{N}$  \cite{Zedini2015}:
\begin{eqnarray}\label{eq:bound_gen}
	\gamma^{\rm MH}_{\gamma_N} = \frac{1}{N}\prod_{i=1}^{N}  C_i^{-\frac{(N-i)}{N}}\gamma_{i}^{\frac{N+1-i}{N}}
\end{eqnarray}

To analyze the statistical performance of the single-hop system in   \eqref{eq:inst_snr} and the multihop system in  \eqref{eq:bound_gen}, we require density functions of the foggy channel, atmospheric turbulence, and pointing errors, which are represented in the following. 

The probability density function (PDF) of the foggy channel is  given as \cite{Esmail2017_Access}:
\begin{flalign}
f_{h_{f_i}}(x) = \frac{z_i^{k_i}}{\Gamma(k_i)}\left(\log\frac{1}{x}\right)^{k_i-1} x^{z_i-1},~0<x\leq 1
\label{eq:pdf_hf}
\end{flalign}
where $z_i=4.343/\beta_i d_i$, $k_i>0$ is the shape parameter and $\beta_i>0$ is the scale parameter. 

Next, we consider the Rician distribution to model non-zero boresight and random jitter of the pointing error  $h_{p_i}$ \cite{Yang2014, Jung2020}
\begin{flalign} 
f_{h_{p_i}}(x) = \frac{\rho_i^2\exp\left(\frac{-s_i^2}{2\sigma_i^2}\right)}{A_{i}^{\rho_i^2}}x^{\rho_i^{2}-1} I_0\left(\frac{s_i}{\sigma_i^2}\sqrt{\frac{w^2_{z_{\rm{eq_i}}}\ln \frac{A_i}{x}}{2}}\right)
\label{eq:pdf_hp}
\end{flalign}
where $0 \leq x\leq A_i$, $w^2_{z_{\rm{eq_i}}}=\frac{w^2_{z_i}\sqrt{A_i\pi}}{2 v_i \exp(-v_i^2)}$, $A_i=[\text{erf}(v_i)]^2$ with $v_i=\sqrt{\frac{r_i^2\pi}{2w^2_{z_i}}}$ as the ratio of aperture radius $r_i$ and beamwidth $w_{z_i}$, and $\rho_i=\frac{w_{z_{\rm{eq_i}}}}{2\sigma_i}$ with $\sigma_i$ as the standard deviation of the jitter and equivalent beamwidth $w_{z_{\rm{eq_i}}}$. Here, $s_i=\sqrt{\mu^2_{x_i}+\mu^2_{y_i}}\neq 0$  models the non-zero boresight, where $\mu_{x_i}\neq 0$ and $\mu_{y_i}\neq 0$ denote the horizontal and vertical displacement between the center of the beam and the center of the detector, respectively. Note that the model in \eqref{eq:pdf_hp} is a generalized one resulting in the special case of zero boresight with $s_i=0$ \cite{Farid2007}.

Finally, we model the atmospheric turbulence channel $h_{t_i}$ using the $\mathcal{F}$-distribution model \cite{Peppas2020_F}:
\begin{flalign}
f_{h_{t_i}}(x)=\frac{a_i^{a_i}(b_i-1)^{b_i}x^{a_i-1}}{\beta(a_i,b_i)(b_i-1)^{a_i+b_i}(B_ix+1)^{a_i+b_i}}
\label{eq:pdf_ht}
\end{flalign}
where $0<h_{t_i}<\infty$, $B_i=\frac{a_i}{(b_i-1)}$, $\beta(\cdot)$ denotes the Beta function, $a_i$, and $b_i$ are the atmospheric refractive-index structure parameters. The atmospheric refractive-index structure parameters $a_i=\frac{1}{\exp(\sigma_{\ln S_i}^2)-1}$ and inner and outer scale parameter of turbulence $b_i=\frac{1}{\exp(\sigma_{\ln L_i}^2)-1}+2$ depends on the small-scale $\sigma_{\ln S_i}^2$ and large-scale $\sigma_{\ln L_i}^2$ log-irradiance variances.

In what follows, we analyze the performance of single hop and multiple hop OWC system using the combined statistical characterization of fog, non-zero boresight pointing errors, and  atmospheric turbulence, as given in  \eqref{eq:pdf_hf}, \eqref{eq:pdf_hp}, and \eqref{eq:pdf_ht}, respectively.

\section{long-range OWC System}
In this section, we  analyze the performance of single-hop and multihop OWC systems under the combined effect of atmospheric turbulence, pointing errors, and fog-induced fading. This is a generalized scenario where the statistical impact of all three channel impairments is considered for a better performance assessment. It is argued that fog and turbulence are inversely correlated \cite{Farid2007}, and thus the presence of one precludes the existence of the other. However, when the density of fog is not high, and the communication range is long, the impact of atmospheric turbulence cannot be ignored.

To proceed with the statistical derivation, first, we find the PDF of ${\cal{F}}$-turbulence channel combined with  the non-zero boresight pointing errors $h_{tp,i}=h_{p_i}h_{t_i}$,  and then include the foggy component to find the resultant PDF for the channel coefficient  $h^{\rm LR}_i=h_{tp,i}h_{f_i}$. Further, we use the 
 series expansion of modified Bessel function $I_0(x)=\sum_{j=0}^{\infty}\frac{\left(\frac{x}{2}\right)^{2j}}{(j!)^2}$ in \eqref{eq:pdf_hp} to get
\begin{flalign}
	&f_{h_{p_i}}(x) = \frac{\rho_i^2\exp\left(\frac{-s_i^2}{2\sigma_i^2}\right)}{A_{i}^{\rho_i^2}}\sum_{j=0}^{\infty}\frac{1}{(j!)^2}\nonumber\\&\left(\frac{s_i^2w^2_{z_{\rm{eq_i}}}}{8\sigma_i^4}\right)^{j} x^{\rho_i^2-1} \left(\ln \frac{A_i}{x}\right)^{j}
	\label{eq:pdf_hp_series}
\end{flalign}
Note that the converging series  expansion in \eqref{eq:pdf_hp_series} facilitates performance analysis in close form.

\subsection{PDF and CDF of SNR}
\begin{my_lemma}\label{pdf_tpf1}
	The PDF and CDF of SNR for a single-hop OWC system with the combined effect of  $\cal{F}$-turbulence channels, fog-induced fading and  generalized pointing errors are given as:
	\begin{flalign}
	&f^{\rm{LR}}_{\gamma_i}(\gamma)=\frac{a_i^{a_i} z_i^{k_i} \rho_i^2\exp\left(\frac{-s_i^2}{2\sigma_i^2}\right)}{2A^{a_i}_i\beta(a_i,b_i)(b_i-1)^{a_i}\sqrt{\gamma\bar{\gamma}_i}} \sum_{j=0}^{\infty}\frac{1}{j!}\nonumber\\&\left(\frac{s_i^2w^2_{z_{\rm{eq_i}}}}{8\sigma_i^4}\right)^{j}\left(\sqrt{\frac{\gamma}{\bar{\gamma}_i}}\right)^{a_i-1}G_{2+j+k_i,2+j+k_i}^{2+j+k_i,1}\nonumber\\&\left[\begin{array}{c} 1-a_i-b_i, \{1+\rho_i^2-a_i\}_0^{j+1}, \{1+z_i-a_i\}_1^{k_i}\\ 0, \{\rho_i^2-a_i\}_0^{j+1}, \{z_i-a_i\}_1^{k_i}\end{array} \left| \frac{B_i}{A_i}\sqrt{\frac{\gamma}{\bar{\gamma}_i}}\right.\right]\nonumber\\&
	\label{combine_pdf_fpt_series}
	\end{flalign}
	
\begin{flalign}
&F^{\rm{LR}}_{\gamma_i}(\gamma)=\frac{a_i^{a_i} z_i^{k_i} \rho_i^2\exp\left(\frac{-s_i^2}{2\sigma_i^2}\right)}{A^{a_i}_i\beta(a_i,b_i)(b_i-1)^{a_i}} \sum_{j=0}^{\infty}\frac{1}{j!}\left(\frac{s_i^2w^2_{z_{\rm{eq_i}}}}{8\sigma_i^4}\right)^{j}\nonumber\\&\left(\sqrt{\frac{\gamma}{\bar{\gamma}_i}}\right)^{a_i}G_{3+j+k_i,3+j+k_i}^{2+j+k_i,2}\nonumber\\&\Bigg[\begin{array}{c} 1-a_i-b_i, 1-a_i, \{1+\rho_i^2-a_i\}_0^{j+1}, \{1+z_i-a_i\}_1^{k_i}\\ 0, \{\rho_i^2-a_i\}_0^{j+1}, \{z_i-a_i\}_1^{k_i}, -a_i \end{array}\nonumber\\& \left| \frac{B_i}{A_i}\sqrt{\frac{\gamma}{\bar{\gamma}_i}}\right.\Bigg]
\label{cdf_fpt_series}
\end{flalign}
\end{my_lemma}

\begin{IEEEproof}
See Appendix A.
\end{IEEEproof}
It can be seen that \eqref{combine_pdf_fpt_series} and \eqref{cdf_fpt_series} generalizes the analysis presented in \cite{Chapala2021} for zero boresight pointing errors. Further, standard functions are available in MATLAB and MATHEMATICA   to compute Meijer's G function.

For ease of presentation, we denote $\phi_i=N+1-i$ and $\psi_i= C_i^{-\frac{(N-i)}{N}}$ in \eqref{eq:bound_gen}, and use the method described in  \cite{bhardwaj2021performance} for terahertz (THz) multihop system to develop 
 the PDF of SNR for $N$-hop  OWC system:
\begin{flalign}
f_{\gamma_N}(\gamma)=\frac{1}{\gamma}\frac{1}{2\pi \jmath}\int_{\cal{L}}^{}\frac{1}{N}\prod_{i=1}^{N}\psi_i\mathbb{E}[\gamma^u_i]\gamma^{-u} du
\label{n_hop_pdf}
\end{flalign}
where $\mathbb{E}[\gamma^u_i]$ is the $u$-th moment of the SNR for the $i$-th hop given by
\begin{flalign}
\mathbb{E}[\gamma^u_i]=\int_{0}^{\infty}\gamma^{u\frac{\phi_i}{N}} f_{\gamma_i}(\gamma)d\gamma
\label{u-th_moment_snr}
\end{flalign}

\begin{my_theorem}	\label{pdf_fpt_series_N}
The PDF and CDF of SNR for a multihop OWC system with the combined effect of  $\cal{F}$-turbulence channels, fog-induced fading and  generalized pointing errors are given as:
	\begin{flalign}
	&f_{\gamma_N}^{\rm LR}(\gamma)=\prod_{i=1}^{N}\frac{\psi_ia_i^{a_i} z_i^{k_i} \rho_i^2\exp\left(\frac{-s_i^2}{2\sigma_i^2}\right)(A_i\sqrt{\bar{\gamma}_i})^{a_i}}{N\gamma B^{2a_i}_i\beta(a_i,b_i)(b_i-1)^{a_i}} \sum_{j=0}^{\infty}\frac{1}{j!}\nonumber\\&\left(\frac{s_i^2w^2_{z_{\rm{eq_i}}}}{8\sigma_i^4}\right)^{j} H_{1+j+k_i,2+j+k_i}^{2+j+k_i,0}\nonumber\\&\Bigg[\begin{array}{c} \{(1+a_i+\rho_i^2, 2\frac{\phi_i}{N})\}_0^{j+1}, \{(1+a_i+z_i, 2\frac{\phi_i}{N})\}_1^{k_i}\\  \{(a_i+\rho_i^2, 2\frac{\phi_i}{N})\}_0^{j+1}, \{(a_i+z_i, 2\frac{\phi_i}{N})\}_1^{k_i}, (2a_i, 2\frac{\phi_i}{N})\end{array} \nonumber\\&\left|\prod_{i=1}^{N} \left(\frac{B_i}{A_i\sqrt{\bar{\gamma}_i}}\right)^{2\frac{\phi_i}{N}}\gamma\right.\Bigg]
	\label{n_hop_pdf_final}
	\end{flalign}
	
	\begin{flalign}
	&F_{\gamma_N}^{\rm {LR}}(\gamma)=\prod_{i=1}^{N}\frac{\psi_ia_i^{a_i} z_i^{k_i} \rho_i^2\exp\left(\frac{-s_i^2}{2\sigma_i^2}\right)(A_i\sqrt{\bar{\gamma}_i})^{a_i}}{N B^{2a_i}_i\beta(a_i,b_i)(b_i-1)^{a_i}} \sum_{j=0}^{\infty}\frac{1}{j!}\nonumber\\&\left(\frac{s_i^2w^2_{z_{\rm{eq_i}}}}{8\sigma_i^4}\right)^{j} H_{2+j+k_i,3+j+k_i}^{2+j+k_i,1}\nonumber\\&\Bigg[\begin{array}{c} (1,1), \{(1+a_i+\rho_i^2, 2\frac{\phi_i}{N})\}_0^{j+1}, \{(1+a_i+z_i, 2\frac{\phi_i}{N})\}_1^{k_i}\\  \{(a_i+\rho_i^2, 2\frac{\phi_i}{N})\}_0^{j+1}, \{(a_i+z_i, 2\frac{\phi_i}{N})\}_1^{k_i}, (2a_i, 2\frac{\phi_i}{N}), (0,1)\end{array} \nonumber\\&\left|\prod_{i=1}^{N} \left(\frac{B_i}{A_i\sqrt{\bar{\gamma}_i}}\right)^{2\frac{\phi_i}{N}}\gamma\right.\Bigg]
	\label{n_hop_cdf_final}
	\end{flalign}	
\end{my_theorem}

\begin{IEEEproof}
See Appendix B.	
\end{IEEEproof}

In the following subsections, we use the above statistical results to derive OP and ABER for the long-range OWC system. 
\subsection{Outage Probability}
Outage probability is defined as the probability that the instantaneous  SNR $\gamma$ falls below a certain threshold SNR $\gamma_{\rm th}$ and is given as
\begin{eqnarray}
	P_{\text{out}}=P(\gamma<\gamma_{\text{th}})=F_{\gamma}(\gamma_{\text{th}})
	\label{eq:outage_prob_general}
\end{eqnarray}
Substituting \eqref{cdf_fpt_series} in \eqref{eq:outage_prob_general}, we can get OP for the single-hop transmissions.
 We can  apply the series expansion of Meijer's G function to derive  the asymptotic expression for the OP in  high SNR regime $\bar{\gamma}_i\to \infty, \forall i$:
 \begin{flalign}
 	&OP_i^{{\rm{LR}},\infty}=\frac{a_i^{a_i} z_i^{k_i} \rho_i^2\exp\left(\frac{-s_i^2}{2\sigma_i^2}\right)}{A^{a_i}_i\beta(a_i,b_i)(b_i-1)^{a_i}} \nonumber\\&\sum_{j=0}^{\infty}\frac{1}{j!}\left(\frac{s_i^2w^2_{z_{\rm{eq_i}}}}{8\sigma_i^4}\right)^{j}\left(\sqrt{\frac{\gamma_{\rm th}}{\bar{\gamma}_i}}\right)^{a_i} \nonumber\\&\sum_{j'=1}^{2+j+k_i}\frac{\prod_{i'=1,i'\neq j'}^{2+j+k_i}\Gamma(\mathcal{V}_{i'}-\mathcal{V}_{j'})\Gamma(a_i+b_i+\mathcal{V}_{j'})\Gamma(a_i+\mathcal{V}_{j'})}{\prod_{i'=3}^{3+j+k_i}\Gamma(\mathcal{U}_{i'}-\mathcal{V}_{j'})\Gamma(1+a_i+\mathcal{V}_{j'})}\nonumber\\&\left(\frac{B_i}{A_i}\sqrt{\frac{\gamma_{\rm th}}{\bar{\gamma}_i}}\right)^{\mathcal{V}_{j'}}\nonumber\\&
 	\label{out_prob_asymp} 
 \end{flalign}
 where $\mathcal{U}_i'=\mathcal{U}_j'=\{1-a_i-b_i, 1-a_i, \{1+\rho_i^2-a_i\}_0^{j+1}, \{1+z_i-a_i\}_1^{k_i}\}$ and $\mathcal{V}_i'=\mathcal{V}_j'=\{0, \{\rho_i^2-a_i\}_0^{j+1}, \{z_i-a_i\}_1^{k_i}, -a_i\}$. Compiling the exponent of $\bar{\gamma}_i=\bar{\gamma}, \forall i$,  the diversity order is derived as $DO_i^{\rm LR}=\min\{\frac{z_i}{2},\frac{\rho_i^2}{2}, \frac{a_i}{2}\}$. Note that the diversity order is independent of the boresight parameters. 

 Similarly, we can substitute the CDF (as given in \eqref{n_hop_cdf_final}) of multihop link in \eqref{eq:outage_prob_general} to get the OP for the short-range multihop transmissions. We can  apply the series expansion of single-varite Fox's H function  \cite[eq. $1.8.4$]{Kilbas} to derive  the asymptotic expression for the OP in  high SNR regime $\bar{\gamma}_i \to \infty, \forall i$:
\begin{eqnarray}
&OP_N^{{\rm{LR}},\infty}=\prod_{i=1}^{N}\frac{\psi_ia_i^{a_i} z_i^{k_i} \rho_i^2\exp\left(\frac{-s_i^2}{2\sigma_i^2}\right)(A_i\sqrt{\bar{\gamma}_i})^{a_i}}{N B^{2a_i}_i\beta(a_i,b_i)(b_i-1)^{a_i}} \sum_{j=0}^{\infty}\frac{1}{j!}\nonumber\\&\left(\frac{s_i^2w^2_{z_{\rm{eq_i}}}}{8\sigma_i^4}\right)^{j}\sum_{j'=1}^{2+j+k_i}\frac{1}{\mathcal{S}_{j'}}\frac{\prod_{i'=1,i'\neq j'}^{2+j+k_i}\Gamma\left(\mathcal{R}_{i'}-\mathcal{R}_{j'}\frac{\mathcal{S}_{i'}}{\mathcal{S}_{j'}}\right)\Gamma\left(\frac{\mathcal{R}_{j'}}{\mathcal{S}_{j'}}\right)}{\prod_{i'=2}^{2+j+k_i}\Gamma\left(\mathcal{P}_{i'}-\mathcal{R}_{j'}\frac{\mathcal{Q}_{i'}}{\mathcal{S}_{j'}}\right)\Gamma\left(1+\frac{\mathcal{R}_{j'}}{\mathcal{S}_{j'}}\right)}\nonumber\\&\left(\prod_{i=1}^{N} \left(\frac{B_i}{A_i\sqrt{\bar{\gamma}_i}}\right)^{2\frac{\phi_i}{N}}\gamma_{\rm th}\right)^{\frac{\mathcal{R}_{j'}}{\mathcal{S}_{j'}}}
\label{out_prob_asymp_N} 
\end{eqnarray}
where $\mathcal{P}_{i'}=\mathcal{P}_{j'}=\{1, \{1+a_i+\rho_i^2\}_0^{j+1}, \{1+a_i+z_i\}_1^{k_i}\}$, $\mathcal{Q}_{i'}=\mathcal{Q}_{j'}=\{1, \{ 2\frac{\phi_i}{N}\}_0^{j+1}, \{ 2\frac{\phi_i}{N}\}_1^{k_i}\}$, $\mathcal{R}_{i'}=\mathcal{R}_{j'}=\{\{a_i+\rho_i^2\}_0^{j+1}, \{a_i+z_i\}_1^{k_i}, 2a_i, 0\}$, and $\mathcal{S}_{i'}=\mathcal{S}_{j'}=\{\{ 2\frac{\phi_i}{N}\}_0^{j+1}, \{2\frac{\phi_i}{N}\}_1^{k_i}, 2\frac{\phi_i}{N}, 1\}$. Similarly, the diversity order is given as $DO_N^{\rm LR}=\sum_{i=1}^N \min \{\frac{z_i}{2},\frac{\rho_i^2}{2}, \frac{a_i}{2}\}$. Comparing the diversity orders for single with the multihop transmissions, the performance scaling with $N$ can be clearly observed. Thus, the use of multiple relays enhances the OWC performance mitigating the effect of pointing errors and path loss due to fog.

\subsection{ABER}
The ABER for a variety of binary and non-binary modulation schemes can be written using the CDF of SNR $(\gamma)$ \cite[Eq. (40)]{Behnam2019} as:
\begin{flalign}
	\bar{P_e}=\frac{q^p}{2\Gamma(p)}\int_{0}^{\infty}\gamma^{p-1}\exp(-q\gamma)F^{\rm {LR}}_{\gamma_N}(\gamma)d\gamma
	\label{eq:ber_3}
\end{flalign}
Substituting \eqref{cdf_fpt_series} in \eqref{eq:ber_3},  applying the definition of Meijer's G-function, and using 	$\int_{0}^{\infty}\gamma^{\frac{2p+a_i+u}{2}-1}\exp(-q\gamma)d\gamma=q^{-\left(\frac{2p+a_i+u}{2}\right)}\Gamma\left(\frac{2p+a_i+u}{2}\right)$, we get the ABER for long-range OWC system for the single hop 
  \begin{flalign}
 	&\bar{P}_{e,i}^{\rm{LR}}=\frac{1}{2\Gamma(p)}\frac{a_i^{a_i} z_i^{k_i} \rho_i^2\exp\left(\frac{-s_i^2}{2\sigma_i^2}\right)}{\left(A_i\sqrt{q\bar{\gamma}_i}\right)^{a_i}\beta(a_i,b_i)(b_i-1)^{a_i}} \nonumber\\&\sum_{j=0}^{\infty}\frac{1}{j!}\left(\frac{s_i^2w^2_{z_{\rm{eq_i}}}}{8\sigma_i^4}\right)^{j}H_{3+j+k_i,4+j+k_i}^{3, 2+j+k_i}\nonumber\\&\Big[\hspace{-0.25cm}\begin{array}{c} (1, 1), \{(1-\rho_i^2+a_i,1)\}_0^{j+1}, \{(1-z_i+a_i, 1)\}_1^{k_i}, (1+a_i, 1)\\ (a_i+b_i,1), (a_i,1), (p+\frac{a_i}{2}, \frac{1}{2}) \{(a_i-\rho_i^2,1)\}_0^{j+1}, \{(a_i-z_i,1)\}_1^{k_i} \end{array} \nonumber\\&\left| \frac{A_i\sqrt{q\bar{\gamma}_i}}{B_i}\right.\Big]
 	\label{eq:ber_lr_df_1}
 \end{flalign}
Similarly, substituting \eqref{n_hop_cdf_final} in \eqref{eq:ber_3} with $\int_{0}^{\infty}\gamma^{p-u-1}\exp(-q\gamma)d\gamma=q^{-p+u}\Gamma(p-u)$, we get the ABER for long-range OWC system for multihop systen
\begin{flalign}
&\bar{P}_{e,N}^{\rm{LR}}=\frac{1}{2\Gamma(p)}\prod_{i=1}^{N}\frac{\psi_ia_i^{a_i} z_i^{k_i} \rho_i^2\exp\left(\frac{-s_i^2}{2\sigma_i^2}\right)(A_i\sqrt{\bar{\gamma}_i})^{a_i}}{N B^{2a_i}_i\beta(a_i,b_i)(b_i-1)^{a_i}} \nonumber\\&\sum_{j=0}^{\infty}\frac{1}{j!}\left(\frac{s_i^2w^2_{z_{\rm{eq_i}}}}{8\sigma_i^4}\right)^{j} H_{3+j+k_i,3+j+k_i}^{2+j+k_i,2}\nonumber\\&\Big[\hspace{-0.25cm}\begin{array}{c} (1,1), (1-p,1), \{(1+a_i+\rho_i^2, 2\frac{\phi_i}{N})\}_0^{j+1}, \{(1+a_i+z_i, 2\frac{\phi_i}{N})\}_1^{k_i}\\  \{(a_i+\rho_i^2, 2\frac{\phi_i}{N})\}_0^{j+1}, \{(a_i+z_i, 2\frac{\phi_i}{N})\}_1^{k_i}, (2a_i, 2\frac{\phi_i}{N}), (0,1)\end{array} \nonumber\\&\left|\prod_{i=1}^{N} \left(\frac{B_i}{A_i\sqrt{\bar{\gamma}_i}}\right)^{2\frac{\phi_i}{N}}\frac{1}{q}\right.\Big]
\label{eq:ber_5}
\end{flalign}
Note that an asymptotic expression for the ABER can be similarly derived by applying the series expansion of single-variate Fox's H function  \cite[eq. $1.8.4$]{Kilbas}, as done for the OP.

\begin{figure*}
	\begin{flalign}
	&OP^{\rm{SR, \infty}}_{N}=\prod_{i=1}^{N}\frac{\psi_iz_i^{k_i}\rho_i^2\exp\left(\frac{-s_i^2}{2\sigma_i^2}\right)}{N m_i^{j+k_i+1}\Gamma(k_i)}(A^2_i\bar{\gamma}_i)^{u\frac{\phi_i}{N}}\sum_{j=0}^{\infty}\frac{1}{(j!)^2}\left(\frac{s_i^2w^2_{z_{\rm{eq_i}}}}{8\sigma_i^4}\right)^{j}\sum_{n=0}^{j}\left(\begin{array}{c} j \\ n \end{array}\right)(-1)^n \Gamma(n+k_i)\Bigg[\Gamma(1+j-n)\sum_{j'=1}^{n-j-1}\nonumber\\&\left(\prod_{i=1}^{N} \frac{\gamma_{\rm th}}{(A^2_i\bar{\gamma}_i)^{\frac{\phi_i}{N}}}\right)^{\frac{\mathcal{C}_{j'}}{\mathcal{D}_{j'}}} \frac{1}{\mathcal{D}_{j'}}\frac{\prod_{i'=1,i'\neq j'}^{n-j-1}\Gamma\left(\mathcal{C}_{i'}-\mathcal{C}_{j'}\frac{\mathcal{D}_{i'}}{\mathcal{D}_{j'}}\right)\Gamma\left(\frac{\mathcal{C}_{j'}}{\mathcal{D}_{j'}}\right)}{\prod_{i'=2}^{n-j}\Gamma\left(\mathcal{A}_{i'}-\mathcal{C}_{j'}\frac{\mathcal{B}_{i'}}{\mathcal{D}_{j'}}\right)\Gamma\left(1+\frac{\mathcal{C}_{j'}}{\mathcal{D}_{j'}}\right)}-\sum_{l=0}^{n+k-1}\frac{1}{l! }\Gamma(1+j-n+l)\sum_{j'=1}^{n-j-l-1}\left(\prod_{i=1}^{N} \frac{\gamma_{\rm th}}{(A^2_i\bar{\gamma}_i)^{\frac{\phi_i}{N}}}\right)^{\frac{\mathcal{R}_{j'}}{\mathcal{S}_{j'}}}\nonumber\\&\frac{1}{\mathcal{S}_{j'}}\frac{\prod_{i'=1,i'\neq j'}^{n-j-l-1}\Gamma\left(\mathcal{R}_{i'}-\mathcal{R}_{j'}\frac{\mathcal{S}_{i'}}{\mathcal{S}_{j'}}\right)\Gamma\left(\frac{\mathcal{R}_{j'}}{\mathcal{S}_{j'}}\right)}{\prod_{i'=2}^{n-j-l}\Gamma\left(\mathcal{P}_{i'}-\mathcal{R}_{j'}\frac{\mathcal{Q}_{i'}}{\mathcal{S}_{j'}}\right)\Gamma\left(1+\frac{\mathcal{R}_{j'}}{\mathcal{S}_{j'}}\right)}\Bigg]
	\label{cout_fp_series_asymp}
	\end{flalign}
	where
	$\mathcal{A}_{i'}=\mathcal{A}_{j'}=\{1, \{1+\frac{\rho_i^2}{m_i}\}_{n=0,j=0}^{n-j-1}\}$, $\mathcal{B}_{i'}=\mathcal{B}_{j'}=\{1,\{\frac{2\phi_i}{m_iN}\}_{n=0,j=0}^{n-j-1}\}$, $\mathcal{C}_{i'}=\mathcal{C}_{j'}=\{\{\frac{\rho_i^2}{m_i}\}_{n=0,j=0}^{n-j-1},0\}$, 
	$\mathcal{D}_{i'}=\mathcal{D}_{j'}=\{\{\frac{2\phi_i}{m_iN}\}_{n=0,j=0}^{n-j-1},1\}$,
	$\mathcal{P}_{i'}=\mathcal{P}_{j'}=\{1,\{2+\frac{\rho_i^2}{m_i}\}_{n=0,j=0,l=0}^{n-j-l-1}\}$,
	$\mathcal{Q}_{i'}=\mathcal{Q}_{j'}=\{1,\{\frac{2\phi_i}{m_iN}\}_{n=0,j=0,l=0}^{n-j-l-1}\}$, 
	$\mathcal{R}_{i'}=\mathcal{R}_{j'}=\{\{1+\frac{\rho_i^2}{m_i}\}_{n=0,j=0,l=0}^{n-j-l-1},0\}$, and
	$\mathcal{S}_{i'}=\mathcal{S}_{j'}=\{\{\frac{2\phi_i}{m_iN}\}_{n=0,j=0,l=0}^{n-j-l-1},1\}$.
	\hrule
\end{figure*}
\section{ short-range OWC System}
The atmospheric turbulence occurs due to the change in the refractive index of the medium when the optical signal is transmitted. It is customary to neglect the atmospheric turbulence for short-range communications, especially in foggy weather conditions. In this section, we analyze the performance of  short-range OWC system over the foggy channel with generalized pointing errors.
\subsection{PDF and CDF of SNR}
\begin{my_lemma}	\label{pdf_fp_series_shop}
	The PDF and CDF of SNR for a single-hop OWC system over fog-induced fading with non-zero boresight pointing errors are given as:
	\begin{flalign}
		&f^{\rm{SR}}_{\gamma_i}(\gamma)=\frac{{z_i}^{k_i}{\rho_i}^2\exp\left(\frac{-{s_i}^2}{2{\sigma_i}^2}\right)}{2\Gamma(k_i)A_{i}^{\rho_i^2}\sqrt{\gamma\bar{\gamma}_i} }\sum_{j=0}^{\infty}\frac{1}{(j!)^2}\left(\frac{{s_i}^2w^2_{z_{\rm{eq_i}}}}{8\sigma_i^4}\right)^{j}\nonumber\\&\left(\sqrt{\frac{\gamma}{\bar{\gamma}_i}}\right)^{\rho_i^2-1}\sum_{n=0}^{j}\left( \begin{array}{c} j \\ n \end{array} \right)(-1)^n\left(\ln\frac{A_i\sqrt{\bar{\gamma}_i}}{\sqrt{\gamma}}\right)^{j-n}\nonumber\\& m^{-n-k_i} \Big[\Gamma(n+k_i)-\Gamma\left(n+k_i,m_i\ln\frac{A_i\sqrt{\bar{\gamma}_i}}{\sqrt{\gamma}}\right)\Big]
		\label{eq:combine_pdf_fp_series}
	\end{flalign}

	\begin{flalign}
		&F^{\rm{SR}}_{\gamma_i}(\gamma)=\frac{{z_i}^{k_i}{\rho_i}^2\exp\left(\frac{-s_i^2}{2\sigma_i^2}\right)}{ \Gamma(k_i)}\sum_{j=0}^{\infty}\frac{1}{(j!)^2}\left(\frac{s_i^2w^2_{z_{\rm{eq_i}}}}{8\sigma_i^4}\right)^{j}\nonumber\\&\sum_{n=0}^{j}\left( \begin{array}{c} j \\ n \end{array} \right)\frac{(-1)^n}{m_i^{j+k_i+1}}\Bigg[\frac{\Gamma(n+k_i)\Gamma\left(1+j-n,\rho_i^2\ln\frac{A_i\sqrt{\bar{\gamma}_i}}{\sqrt{\gamma}}\right)}{\left(\frac{\rho_i^2}{m_i}\right)^{j-n+1}}\nonumber\\&-\frac{(n+k_i-1)!\Gamma\left(1+j-n+l,\left(\frac{\rho_i^2}{m_i}+1\right)m_i\ln\frac{A_i\sqrt{\bar{\gamma}_i}}{\sqrt{\gamma}}\right)}{\left(\frac{\rho_i^2}{m_i}+1\right)^{j-n+l+1}}\Bigg]
		\label{eq:combine_cdf_fp_series}
	\end{flalign}
	where $0\leq \gamma\leq A_i^2 \bar{\gamma}_i$.
\end{my_lemma}

\begin{IEEEproof}
	See 	Appendix C.
\end{IEEEproof}

It can be seen that \eqref{eq:combine_pdf_fp_series} and \eqref{eq:combine_cdf_fp_series} generalizes the analysis presented in \cite{Esmail2017_Photonics} for zero boresight pointing errors. Thus, substituting $s=0$ in \eqref{eq:combine_pdf_fp_series} and \eqref{eq:combine_cdf_fp_series}, we can get PDF and CDF for the zero boresight OWC system, respectively. 

Similar to the previous subsection, We use \eqref{n_hop_pdf} to develop statistical results for the short-range multihop transmissions.

\begin{my_theorem}\label{pdf_fp_series_Nhop}
	The PDF and CDF of SNR for a multihop OWC system over fog-induced fading with non-zero boresight pointing errors are given as:
\end{my_theorem}
\begin{flalign}
	&f^{\rm{SR}}_{\gamma_N}(\gamma)=\prod_{i=1}^{N}\frac{\psi_iz_i^{k_i}\rho_i^2\exp\left(\frac{-s_i^2}{2\sigma_i^2}\right)}{N\gamma m_i^{j+k_i+1}\Gamma(k_i)}\sum_{j=0}^{\infty}\frac{1}{(j!)^2}\times\nonumber\\&\left(\frac{s_i^2w^2_{z_{\rm{eq_i}}}}{8\sigma_i^4}\right)^{j}\sum_{n=0}^{j}\left( \begin{array}{c} j \\ n \end{array} \right)(-1)^n \Big[\Gamma(n+k_i)\Gamma(1+j-n)\times\nonumber\\&H_{n-j-1,n-j-1}^{n-j-1,0}\left[\begin{array}{c} \{(1+\frac{\rho_i^2}{m_i},\frac{2\phi_i}{m_iN})\}_{n=0,j=0}^{n-j-1}\\  \{(\frac{\rho_i^2}{m_i},\frac{2\phi_i}{m_iN})\}_{n=0,j=0}^{n-j-1}\end{array} \left|\prod_{i=1}^{N} \frac{\gamma}{(A^2_i\bar{\gamma}_i)^{\frac{\phi_i}{N}}}\right.\right]\nonumber\\&-(n+k-1)!\nonumber\\&\times\sum_{l=0}^{n+k-1}\frac{1}{l! }\Gamma(1+j-n+l)H_{n-j-l-1,n-j-l-1}^{n-j-l-1,0}\nonumber\\&\left[\begin{array}{c} \{(2+\frac{\rho_i^2}{m_i},\frac{2\phi_i}{m_iN})\}_{n=0,j=0,l=0}^{n-j-l-1}\\  \{(1+\frac{\rho_i^2}{m_i},\frac{2\phi_i}{m_iN})\}_{n=0,j=0,l=0}^{n-j-l-1}\end{array} \left|\prod_{i=1}^{N} \frac{\gamma}{(A^2_i\bar{\gamma}_i)^{\frac{\phi_i}{N}}}\right.\right]\Big]
	\label{combine_pdf_fp_series_final}
\end{flalign}

\begin{flalign}
	&F^{\rm{SR}}_{\gamma_N}(\gamma)=\prod_{i=1}^{N}\frac{\psi_iz_i^{k_i}\rho_i^2\exp\left(\frac{-s_i^2}{2\sigma_i^2}\right)}{N m_i^{j+k_i+1}\Gamma(k_i)}\sum_{j=0}^{\infty}\frac{1}{(j!)^2}\times\nonumber\\&\left(\frac{s_i^2w^2_{z_{\rm{eq_i}}}}{8\sigma_i^4}\right)^{j}\sum_{n=0}^{j}\left(\begin{array}{c} j \\ n \end{array}\right)(-1)^n \Big[\Gamma(n+k_i)\Gamma(1+j-n)\times\nonumber\\&H_{n-j,n-j}^{n-j-1,1}\left[\begin{array}{c} (1,1), \{(1+\frac{\rho_i^2}{m_i},\frac{2\phi_i}{m_iN})\}_{n=0,j=0}^{n-j-1}\\  \{(\frac{\rho_i^2}{m_i},\frac{2\phi_i}{m_iN})\}_{n=0,j=0}^{n-j-1}, (0,1)\end{array} \left|\prod_{i=1}^{N} \frac{\gamma}{(A^2_i\bar{\gamma}_i)^{\frac{\phi_i}{N}}}\right.\right]\nonumber\\&-(n+k-1)!\sum_{l=0}^{n+k-1}\frac{1}{l! }\Gamma(1+j-n+l)H_{n-j-l,n-j-l}^{n-j-l-1,1}\nonumber\\&\left[\begin{array}{c} (1,1), \{(2+\frac{\rho_i^2}{m_i},\frac{2\phi_i}{m_iN})\}_{n=0,j=0,l=0}^{n-j-l-1}\\  \{(1+\frac{\rho_i^2}{m_i},\frac{2\phi_i}{m_iN})\}_{n=0,j=0,l=0}^{n-j-l-1}, (0,1)\end{array} \left|\prod_{i=1}^{N} \frac{\gamma}{(A^2_i\bar{\gamma}_i)^{\frac{\phi_i}{N}}}\right.\right]\Big]
	\label{combine_cdf_fp_series_new}
\end{flalign}

\begin{IEEEproof}
	See 	Appendix D.
\end{IEEEproof}

\subsection{Outage Probability}
Substituting \eqref{eq:combine_cdf_fp_series} in \eqref{eq:outage_prob_general}, we can get an exact OP for the single-hop transmissions.
Further, we apply $\frac{\Gamma(s,x)}{x^{s-1}\exp(-x)}\rightarrow 1$ as $x \rightarrow \infty$ in  \eqref{eq:combine_cdf_fp_series} to find the asymptotic expression of the OP at high SNR
\begin{flalign}
	&OP^{\rm{SR, \infty}}_{i}=\frac{{z_i}^{k_i}{\rho_i}^2\exp\left(\frac{-s_i^2}{2\sigma_i^2}\right)}{ \Gamma(k_i)}\sum_{j=0}^{\infty}\frac{1}{(j!)^2}\left(\frac{s_i^2w^2_{z_{\rm{eq_i}}}}{8\sigma_i^4}\right)^{j}\nonumber\\&\sum_{n=0}^{j}\left( \begin{array}{c} j \\ n \end{array} \right)\frac{(-1)^n\Gamma(n+k_i)}{m_i^{j+k_i+1}}\Bigg[\frac{\left(\rho_i^2\ln\frac{A_i\sqrt{\bar{\gamma}_i}}{\sqrt{\gamma_{\rm th}}}\right)^{j-n}\left(\frac{A_i\sqrt{\bar{\gamma}_i}}{\sqrt{\gamma_{\rm th}}}\right)^{-\rho_i^2}}{\left(\frac{\rho_i^2}{m_i}\right)^{j-n+1}}\nonumber\\&-\frac{\left(z_i\ln\frac{A_i\sqrt{\bar{\gamma}_i}}{\sqrt{\gamma_{\rm th}}}\right)^{j-n+l}\left(\frac{A_i\sqrt{\bar{\gamma}_i}}{\sqrt{\gamma_{\rm th}}}\right)^{-z_i}}{\left(\frac{z_i}{m_i}\right)^{j-n+l+1}}\Bigg]
	\label{eq:combine_cdf_fp_series_asymp}
\end{flalign}
Compiling the exponent of $\bar{\gamma}_i$, the diversity order for the single-hop transmission is  $DO_i^{\rm SR}=\min\{\frac{z_i}{2},\frac{\rho_i^2}{2}\}$. It can be seen that the diversity order for the single-hop OWC system is exactly the same as that of the OWC, with zero boresight pointing errors \cite{Rahman2020_Systems}. Thus, there is no impact of non-zero boresight parameter $s\neq 0$ on the diversity order of the system.

Similarly, we can substitute the CDF (as given in \eqref{combine_cdf_fp_series_new}) of multihop link in \eqref{eq:outage_prob_general} to get the OP for the short-range multihop transmissions. We can  apply the series expansion of single-variate Fox's H function  \cite[eq. $1.8.4$]{Kilbas} to derive  the asymptotic expression for the OP in  high SNR regime $\bar{\gamma}_i\to \infty, \forall i$ in \eqref{cout_fp_series_asymp}.

Using \eqref{cout_fp_series_asymp}, we can find the diversity order  as  $DO_N^{\rm LR}=\sum_i^N\ \min \{\frac{z_i}{2},\frac{\rho_i^2}{2}\}$, which is a special case for the long-range multihop communications. Thus, the use of multiple relays provides the capability to mitigate the effect of pointing errors and signal attenuation due to the fog.

\subsection{ABER}
The ABER for a variety of binary modulation schemes can be written using the PDF of SNR as
\begin{flalign}
	\bar{P_e}=p\int_{0}^{\infty}Q(\sqrt{q\gamma})f^{}_{\gamma_i}(\gamma)d\gamma
	\label{eq:ber_1_pdf}
\end{flalign}
where   the constants $p$, $q$, provide flexibility to design various modulation schemes.
Using \eqref{eq:combine_pdf_fp_series} in	\eqref{eq:ber_1_pdf} with the series expansion $\Gamma(a,t) \triangleq (a-1)! e^{-t} \sum_{m=0}^{a-1} \frac{t^m}{m!}$, we get
\begin{flalign}
	&\bar{P}_{e,i}^{\rm SR}=p\int_{0}^{A^2_i\bar{\gamma}_i}Q(\sqrt{q\gamma})\frac{z_i^{k_i}\rho_i^2\exp\left(\frac{-s_i^2}{2\sigma_i^2}\right)}{2\Gamma(k_i)A_{0}^{\rho_i^2}\sqrt{\gamma\bar{\gamma}_i} }\sum_{j=0}^{\infty}\frac{1}{(j!)^2}\nonumber\\&\left(\frac{s_i^2w^2_{z_{\rm{eq_i}}}}{8\sigma_i^4}\right)^{j}\left(\sqrt{\frac{\gamma}{\bar{\gamma}_i}}\right)^{\rho_i^2-1}\sum_{n=0}^{j}\left( \begin{array}{c} j \\ n \end{array} \right)(-1)^n\nonumber\\&\left(\ln\frac{A_i\sqrt{\bar{\gamma}_i}}{\sqrt{\gamma}}\right)^{j-n}m_i^{-n-k_i} \Bigg[\Gamma(n+k_i)-(n+k_i-1)!\nonumber\\&\left(\frac{A_i\sqrt{\bar{\gamma}_i}}{\sqrt{\gamma}}\right)^{-m_i}\sum_{l=0}^{n+k_i-1}\frac{\left(m_i\ln\frac{A_i\sqrt{\bar{\gamma}_i}}{\sqrt{\gamma}}\right)^l}{l!}\Bigg]d\gamma
	\label{combine_pe_fp_series_sr}
\end{flalign}
Applying $Q(\sqrt{q\gamma})=G_{1,2}^{2,0}\left[\begin{array}{c} 1\\  0,\frac{1}{2}\end{array} \left|\frac{q}{2}\gamma\right.\right]=\frac{1}{2\pi i}\int_{L}^{}\frac{\Gamma(-u)\Gamma(\frac{1}{2}-u)}{\Gamma(1-u)}\left(\frac{q}{2}\gamma\right)^u$ and substituting $\ln\frac{A_i\sqrt{\bar{\gamma}_i}}{\sqrt{\gamma}}=t$ and $d\gamma=-2A^2_i\bar{\gamma}_i\exp(-2t)dt$ in \eqref{combine_pe_fp_series_sr}, we solve the integrals	$\int_{0}^{\infty}t^{j-n} \exp[-(\rho^2_i+u)t]dt=\Gamma(1+j-n)\left(\frac{\Gamma(\rho^2_i+2u)}{\Gamma(1+\rho^2_i+2u)}\right)^{n-j-1}$ and 
$\int_{0}^{\infty}t^{j-n+l} \exp[-(z_i+u)t]dt=\Gamma(1+j-n+l)
\left(\frac{\Gamma(z_i+2u)}{\Gamma(1+z_i+2u)}\right)^{n-j-l-1}$. Using these integral, we apply the definition of Fox's H-function in \eqref{combine_pe_fp_series_sr} to get a closed-form expression for the ABER for the single-hop OWC system as:
\begin{flalign}
	&\bar{P}_{e,i}^{\rm{SR}}=\frac{z_i^{k_i}\rho_i^2\exp\left(\frac{-s_i^2}{2\sigma_i^2}\right)}{2\Gamma(k_i)A_{0}^{\rho_i^2}\sqrt{\gamma\bar{\gamma}_i} }\sum_{j=0}^{\infty}\frac{1}{(j!)^2}\left(\frac{s_i^2w^2_{z_{\rm{eq_i}}}}{8\sigma_i^4}\right)^{j}\nonumber\\&\left(\sqrt{\frac{\gamma}{\bar{\gamma}_i}}\right)^{\rho_i^2-1}\sum_{n=0}^{j}\left( \begin{array}{c} j \\ n \end{array} \right)(-1)^n\left(\ln\frac{A_i\sqrt{\bar{\gamma}_i}}{\sqrt{\gamma}}\right)^{j-n}m_i^{-n-k_i} \nonumber\\&\Bigg[\Gamma(n+k_i)H_{n-j+1,n-j}^{n-j-1,2} \Bigg[\begin{array}{c} (1,1), (\frac{1}{2}, 1), \{(1+\rho^2_i,2)\}^{n-j-1}\\\{(\rho^2_i,2)\}^{n-j-1}, (0,1)\end{array} \nonumber\\&\left| \frac{2}{qA^2_i\bar{\gamma}_i}\right.\Bigg]-(n+k_i-1)!\nonumber\\&\sum_{l=0}^{n+k_i-1}\frac{\Gamma(1+j-n+l)}{l!}H_{n-j-l+1,n-j-l}^{n-j-l-1,2} \nonumber\\&\left[\begin{array}{c} (1,1), (\frac{1}{2}, 1), \{(1+z_i,2)\}^{n-j-l-1}\\\{(z_i,2)\}^{n-j-l-1}, (0,1)\end{array} \left| \frac{2}{qA^2_i\bar{\gamma}_i}\right.\right]\Bigg]
	\label{combine_ber_fp_series_sr1}
\end{flalign} 

\begin{figure*}[h]
	\begin{flalign}
	&\bar{P}_{e,N}^{\rm SR}=p\prod_{i=1}^{N}\frac{\psi_iz_i^{k_i}\rho_i^2\exp\left(\frac{-s_i^2}{2\sigma_i^2}\right)}{N m_i^{j+k_i+1}\Gamma(k_i)}\sum_{j=0}^{\infty}\frac{1}{(j!)^2}\left(\frac{s_i^2w^2_{z_{\rm{eq_i}}}}{8\sigma_i^4}\right)^{j}\sum_{n=0}^{j}\left( \begin{array}{c} j \\ n \end{array} \right)(-1)^n \Big[\Gamma(n+k_i)\Gamma(1+j-n)\nonumber\\&H_{1,0:1,2;n-j-1,n-j-1}^{0,1:2,0;n-j-1,0}\left[\begin{array}{c} (1,1):(1,1);\{(1+\frac{\rho_i^2}{m_i},\frac{2\phi_i}{m_iN})\}_{n=0,j=0}^{n-j-1}\\  (0,1):(0,1),(\frac{1}{2},1);\{(\frac{\rho_i^2}{m_i},\frac{2\phi_i}{m_iN})\}_{n=0,j=0}^{n-j-1}\end{array} \left|\prod_{i=1}^{N}\frac{A_i^2\bar{\gamma}_iq}{2},\prod_{i=1}^{N} (A^2_i\bar{\gamma}_i)^{1-\frac{\phi_i}{N}}\right.\right]-(n+k-1)!\sum_{l=0}^{n+k-1}\frac{1}{l! }\nonumber\\&\Gamma(1+j-n+l)H_{1,0:1,2;n-j-l-1,n-j-l-1}^{0,1:2,0;n-j-l-1,0}\left[\begin{array}{c} (1,1):(1,1);\{(2+\frac{\rho_i^2}{m_i},\frac{2\phi_i}{m_iN})\}_{n=0,j=0,l=0}^{n-j-l-1}\\  (0,1):(0,1),(\frac{1}{2},1);\{(1+\frac{\rho_i^2}{m_i},\frac{2\phi_i}{m_iN})\}_{n=0,j=0,l=0}^{n-j-l-1}\end{array} \left|\prod_{i=1}^{N}\frac{A_i^2\bar{\gamma}_iq}{2},\prod_{i=1}^{N} (A^2_i\bar{\gamma}_i)^{1-\frac{\phi_i}{N}}\right.\right]\Big]
	\label{eq:ber_2_sr_af3}
	\end{flalign}
	\hrule
\end{figure*}

To derive the ABER for the multihop system, we substitute \eqref{combine_pdf_fp_series_final} in  \eqref{eq:ber_1_pdf}, apply $Q(\sqrt{q\gamma})=G_{1,2}^{2,0}\left[\begin{array}{c} 1\\  0,\frac{1}{2}\end{array} \left|\frac{q}{2}\gamma\right.\right]=\frac{1}{2\pi \jmath}\int_{\cal{L}}^{}\frac{\Gamma(-u_1)\Gamma(\frac{1}{2}-u_1)}{\Gamma(1-u_1)}\left(\frac{q}{2}\gamma\right)^{u_1}$ and definition of Fox's H-function, we get

	\begin{flalign}
	&\bar{P}_{e,N}^{\rm SR}=p\int_{0}^{A^2_i\bar{\gamma}_i}\frac{1}{2\pi \jmath}\int_{\cal{L}}^{}\frac{\Gamma(-u_1)\Gamma(\frac{1}{2}-u_1)}{\Gamma(1-u_1)}
	\left(\frac{q}{2}\gamma\right)^{u_1}\prod_{i=1}^{N}\nonumber\\&\frac{\psi_iz_i^{k_i}\rho_i^2
		\exp\left(\frac{-s_i^2}{2\sigma_i^2}\right)}{N\gamma m_i^{j+k_i+1}\Gamma(k_i)}\sum_{j=0}^{\infty}\frac{1}{(j!)^2}\left(\frac{s_i^2w^2_{z_{\rm{eq_i}}}}{8\sigma_i^4}\right)^{j}\nonumber\\&\sum_{n=0}^{j}\left( \begin{array}{c} j \\ n \end{array} \right)(-1)^n \Big[\Gamma(n+k_i)\Gamma(1+j-n)\nonumber\\&\frac{1}{2\pi \jmath}\int_{\cal{L}}^{}\left(\frac{\Gamma\left(\frac{\rho_i^2}{m_i}+\frac{2\phi_i}{m_iN}u_2\right)}{\Gamma\left(1+\frac{\rho_i^2}{m_i}+\frac{2\phi_i}{m_iN}u_2\right)}\right)^{n-j-1}\left(\frac{\gamma}{(A^2_i\bar{\gamma}_i)^{\frac{\phi_i}{N}}}\right)^{-u_2}\nonumber\\&-(n+k-1)!
	\sum_{l=0}^{n+k-1}\frac{1}{l! }\Gamma(1+j-n+l)\nonumber\\&\frac{1}{2\pi \jmath}\int_{\cal{L}}^{}\Bigg(\frac{\Gamma(1+\frac{\rho_i^2}{m_i}+\frac{2\phi_i}{m_iN}u_2)}{\Gamma(2+\frac{\rho_i^2}{m_i}+\frac{2\phi_i}{m_iN}u_2)}\Bigg)^{n-j-l-1}\bigg(\frac{\gamma}{(A^2_i\bar{\gamma}_i)^{\frac{\phi_i}{N}}}\bigg)^{-u_2}\Big]d\gamma
	\label{eq:ber_2_sr_af1}
	\end{flalign}

Solving the above inner integral $\int_{0}^{A^2_i\bar{\gamma}_i}\gamma^{u_1-u_2-1}d\gamma=\frac{(A^2_i\bar{\gamma}_i)^{u_1-u_2}}{u_1-u_2}=\frac{\Gamma(u_1-u_2)}{\Gamma(1+u_1-u_2)}$ and substituting $u_2\rightarrow-u_2$, and  applying the definition of bivariate Fox's H-function \cite{Mittal_1972}, we get the ABER in \eqref{eq:ber_2_sr_af3}.

\begin{table}[tp]	
	\renewcommand{\arraystretch}{01}
	\caption{Simulation Parameters}
	\label{table:simulation_parameters}
	\centering
	\begin{tabular}{|c|p{1.15cm}|p{2.3cm}|}
		\hline 	
		Transmitted power &$P_t$ & $0$ to $40$ \mbox{dBm} \\ \hline
		Responsitivity &	$R$ & $0.4$ \mbox{A/W}\\ \hline
		
		AWGN variance &$\sigma_w^2$ & $10^{-14}~\rm {A^2/GHz}$ \\ \hline	
		Long-range distance &$d$ &  $1500$ \mbox{m}\\ \hline	
			Short-range distance &$d$  & $1000$ \mbox{m} \\ \hline	
		Shape parameter of fog &$k$ & \{2.32, 5.49, 6.00\} \\ \hline	
		Scale parameter of fog &$\beta$ & \{13.12, 12.06, 23.00\}\\ \hline Aperture diameter & $D=2a_r$ & $10$ \mbox{cm} \\ \hline Normalized beam-width &$w_z/a_r$ & \{3, 6\}\\ \hline Jitter standard deviation &$\sigma_{s}$ & \{5-20\} ${\rm{cm}}$ \\ \hline Horizontal and vertical displacement &$\mu_{x_i}, \mu_{y_i}$ & $\{0.1, 0.1\}$ \\ \hline Wavelength &$\lambda$ & $1550$ \mbox{nm} \\ \hline  Turbulence parameters &$a_i$,$b_i$  & \{4.5916, 2.3378, 1.4321\}, \{7.0941, 4.5323, 3.4948\}\\ \hline
	\end{tabular}	 
\end{table}

\begin{figure*}[tp]
	\subfigure[OP for different turbulence conditions over light fog.]{\includegraphics[scale= 0.30] {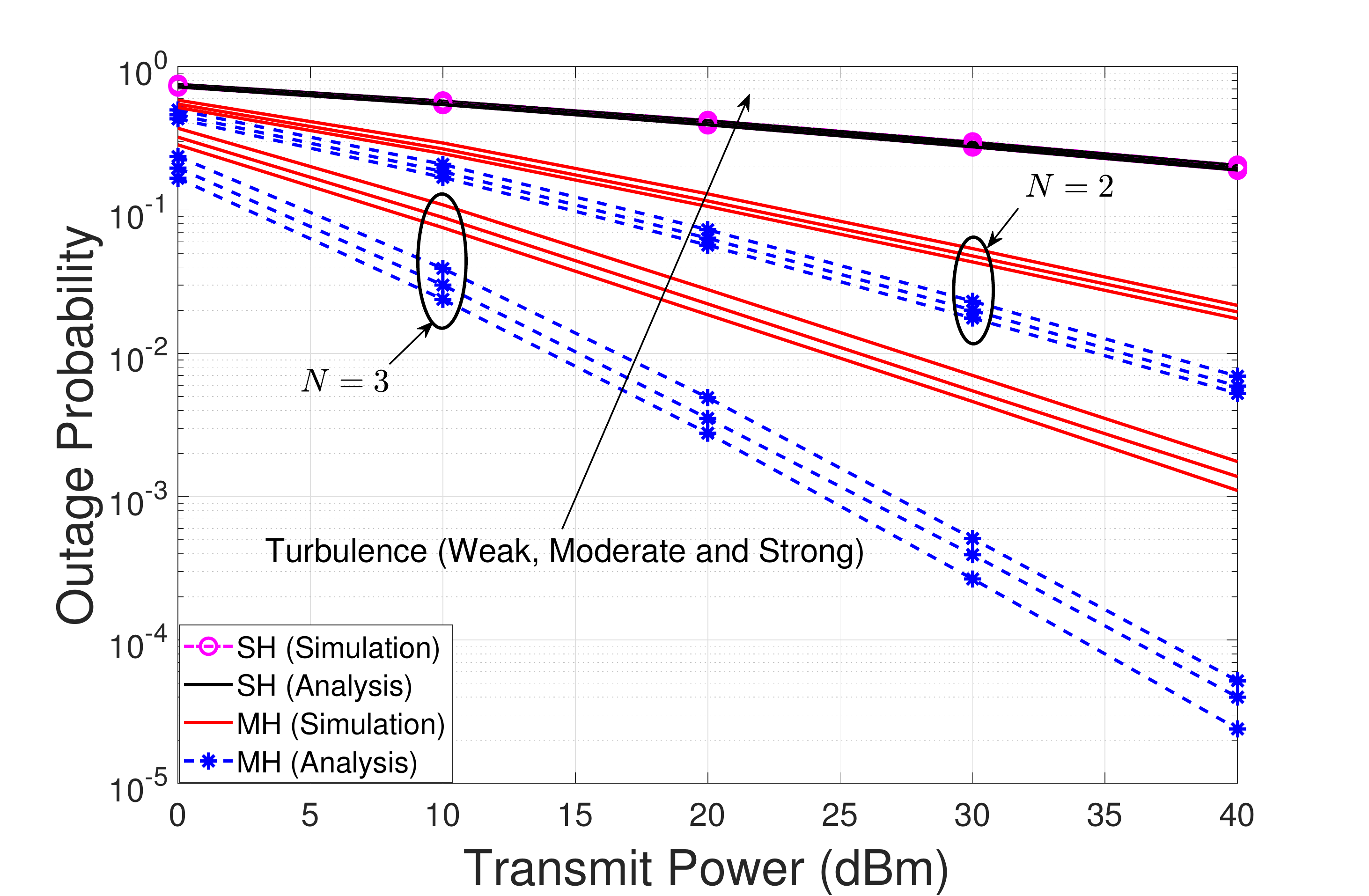}}
	\subfigure[ABER over different fog and turbulence conditions.]{\includegraphics[scale= 0.30] {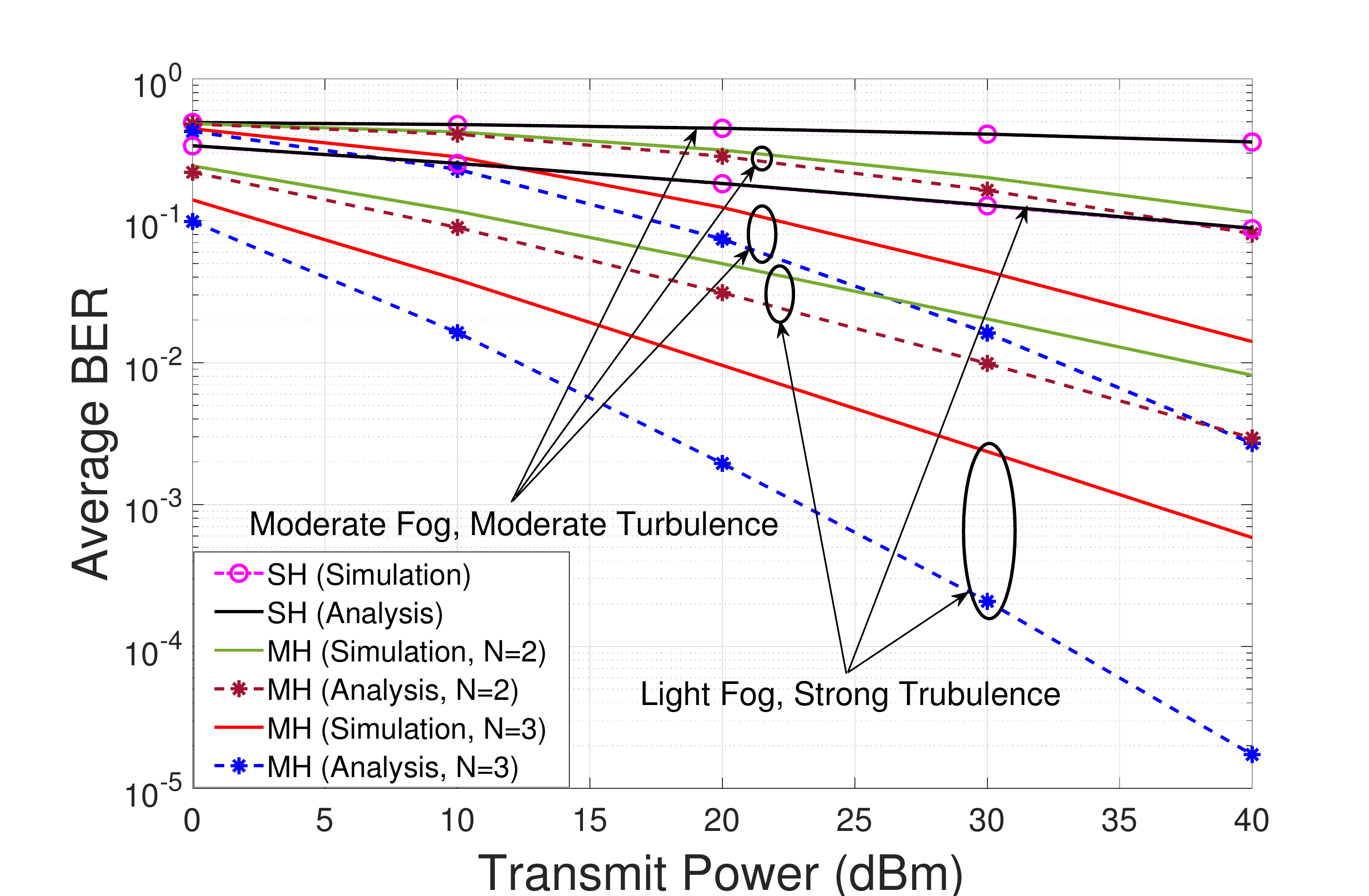}}	
	\centering	\subfigure[ABER over light fog and different turbulence conditions at transmit power $P_i=10{\rm{dBm}}$.]{\includegraphics[scale= 0.30] {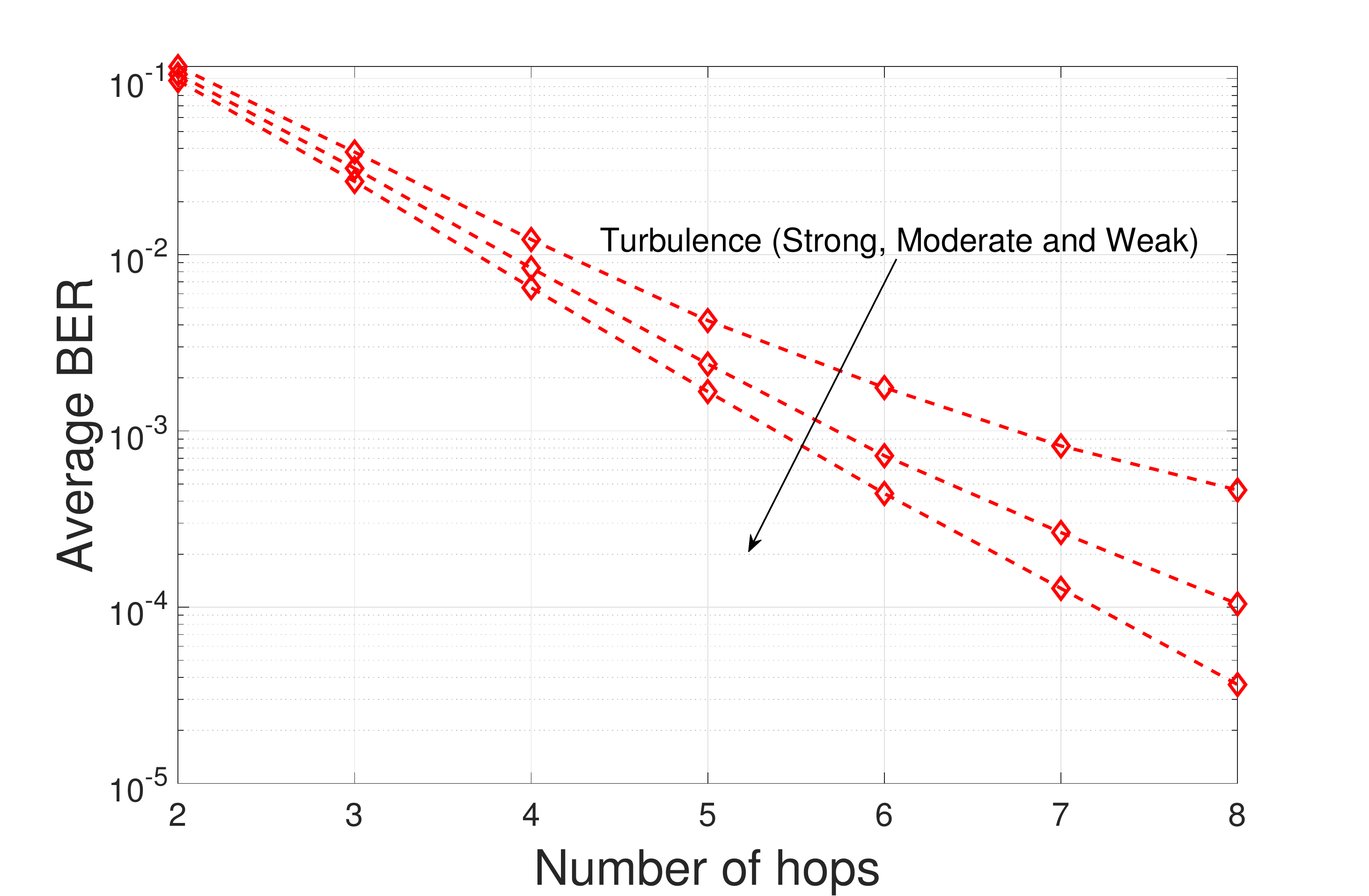}}	
	
	\caption{Performance for long-range OWC systems over random fog, atmospheric turbulence  with pointing error parameters $\frac{w_z}{a_r}=6$ and $\sigma_s=5~{\rm{cm}}$ .}
	\label{out_prob_fpt}
\end{figure*}

\begin{figure*}[tp]
	\subfigure[OP analysis over light fog with different pointing errors conditions.]{\includegraphics[scale= 0.30] {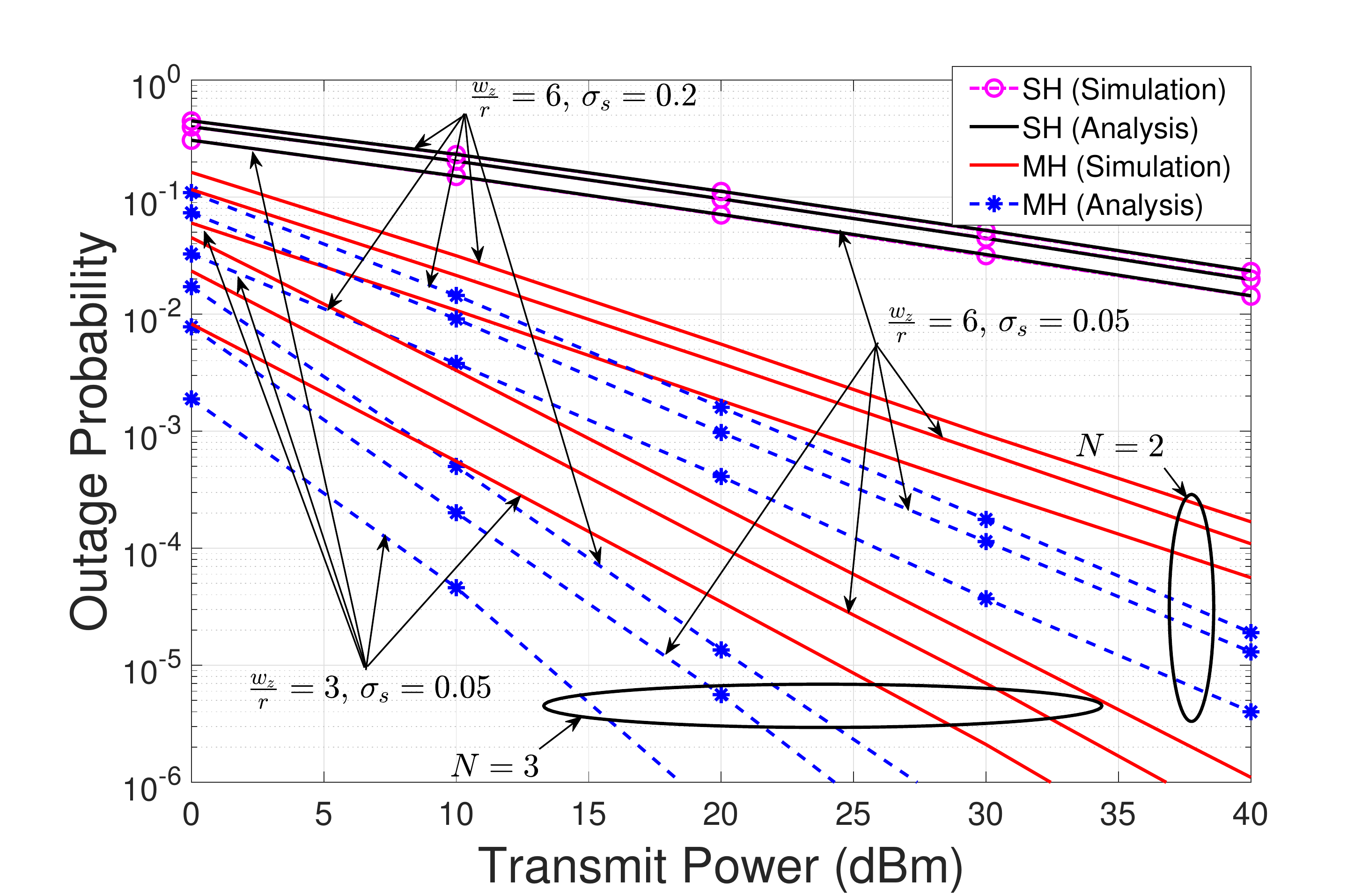}}
	\subfigure[OP analysis over light fog and moderate fog with different pointing errors conditions at transmit power $P_i=10{\rm{dBm}}$.]{\includegraphics[scale= 0.30] {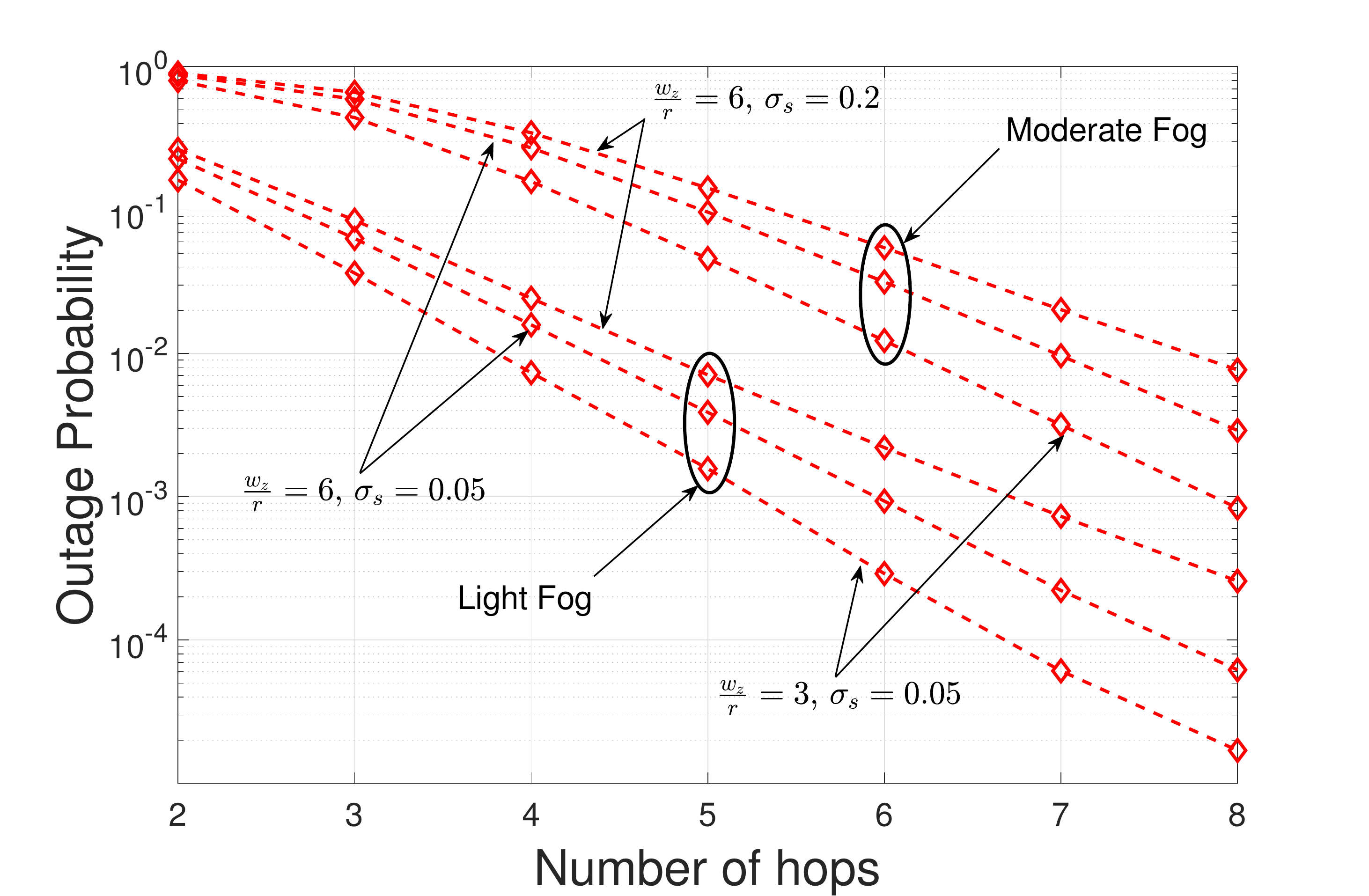}}	
	\centering	\subfigure[ABER over moderate fog with pointing error parameters $\frac{w_z}{a_r}=6$ and $\sigma_s=5~{\rm{cm}}$.]{\includegraphics[scale= 0.30] {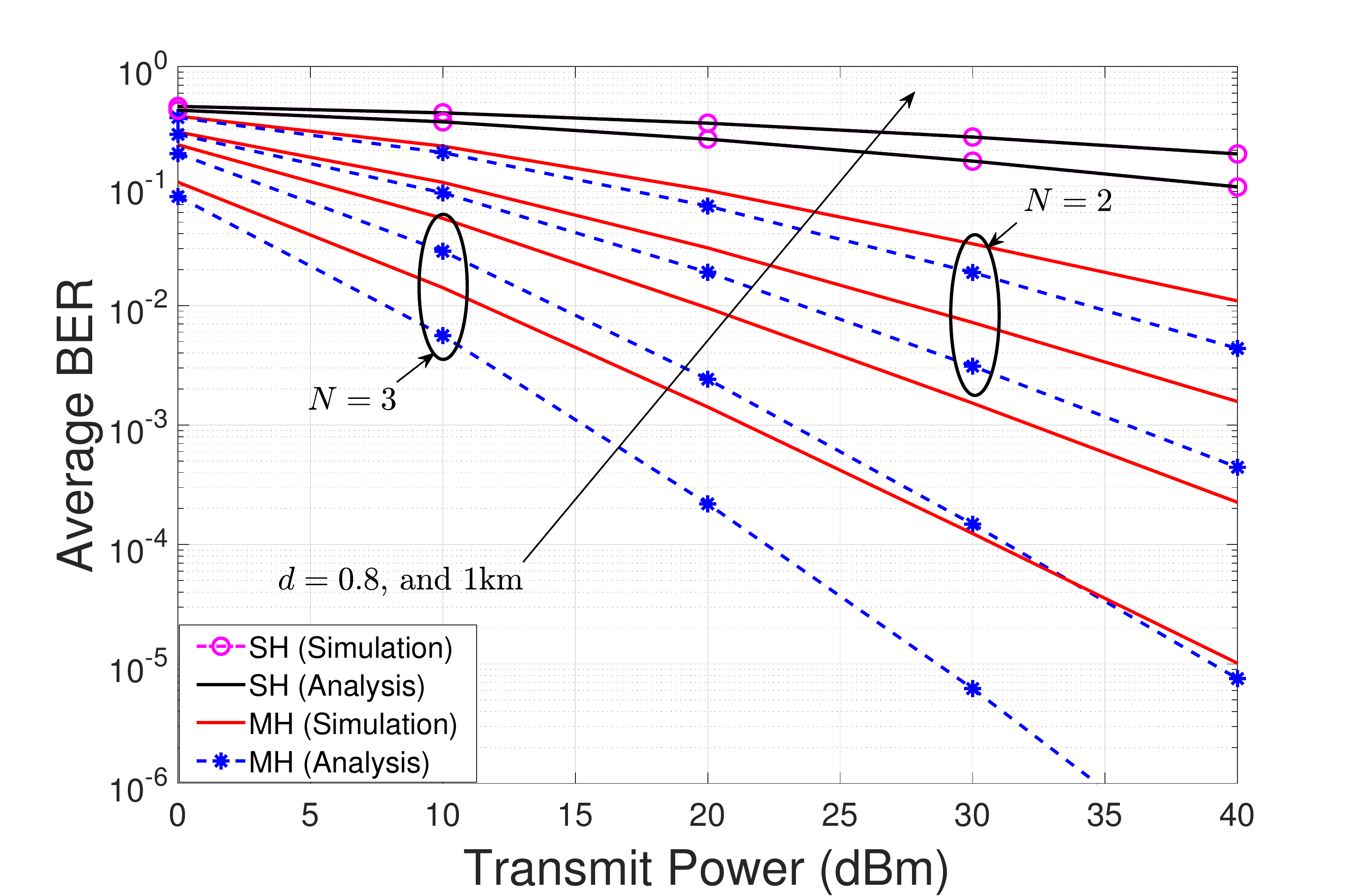}}	
	
	\caption{Performance for short-range OWC systems over random fog and pointing errors.}
	\label{out_prob_fp}
\end{figure*}

\section{Simulation and Numerical Analysis}

In this section, we investigate the performance of multihop OWC system over atmospheric turbulence and fog with generalized pointing errors using both numerical and Monte Carlo (MC) simulation (averaged over $10^7$ channel realizations) approaches. We use $20$ terms of the series expansion to numerically evaluate the derived expressions. We consider two simulation scenarios: short-range (link distance limited to $1000$ \mbox{m}) communication over fog with generalized pointing errors and long-range (link distance above $1000$ \mbox{m}) communication over atmospheric turbulence and fog with generalized pointing errors. We  validate our derived analytical expressions with numerical and simulation results.  We  demonstrate the significance of multihop relaying  as compared with the single-hop transmission for various link distances and channel impairments (turbulence, fog, and pointing errors).  We list the simulation parameters in Table \ref{table:simulation_parameters}.

In Fig.~\ref{out_prob_fpt}, we demonstrate the significance of multihop relaying  for the OWC system under the combined effect of atmospheric turbulence, pointing errors, and random fog.  We consider  different atmospheric turbulence conditions (weak, moderate, and strong) with light fog and compare the OP with the single-hop, two hops, and three hops OWC system, as depicted in Fig.~\ref{out_prob_fpt}(a). The impact of turbulence intensity is not significant if the performance degradation with strong turbulence is compared with the weak one. Further, multihop relaying with $N=3$ significantly improves the OP.   

We take two scenarios where the inverse correlation between turbulence and fog has been adopted by considering strong turbulence with light fog and moderate turbulence in the high-density moderate fog condition. We plot the ABER performance of the direct link, two hop, and three hop OWC system, as shown in  Fig.~\ref{out_prob_fpt}(b). The  ABER performance is extremely poor  for both moderate and light foggy conditions with the direct transmission for the link distance of $1500$\mbox{m}.  However,  the ABER is significantly improved  when the multihop relaying is employed. A dual-hop transmission is sufficient to achieve an acceptable ABER of $10^{-3}$ at a transmit power in light fog, where a few additional hops  are required to achieve the same performance in moderate fog conditions.  In Fig.~\ref{out_prob_fpt}(c), we describe the impact of the number of hops on the ABER performance for different turbulence conditions (weak, moderate, and strong). The figure shows the performance scaling  with the number of hops achieving an acceptable ABER of $10^{-3}$ within $5$ hops at a $10$\mbox{dBm} transmit power.  

In Fig.~\ref{out_prob_fp}, we demonstrate the performance of OWC for short-range communication under the combined effect of random fog and pointing errors with negligible atmospheric turbulence. We plot the OP of the OWC system for short-range communication, as depicted in Fig.~\ref{out_prob_fp}(a).  We demonstrate the impact of fog density and pointing errors on the performance of multihop OWC system by considering a single hop (direct link), two hops, and three hops scenarios. It can be seen from Fig.~\ref{out_prob_fp}(a) that there is a significant improvement in the OP performance with an  increase in the number of hops. The figure shows that single hop transmission is not useful even at a transmit power of $40$\mbox{dBm} with OP more than $10^{-2}$. The dual-hop and triple-hop transmissions attain an acceptable  OP  in the range of $10^{-4}$ and $10^{-6}$, respectively. We can also observe the  impact of pointing errors parameters ($\frac{w_z}{a_r}=\{3, 6\}$ and $\sigma_s=\{5-20\}~{\rm{cm}}$) on the OP. 
In Fig.~\ref{out_prob_fp}(b), we plot the OP versus the number of hops at a transmit power $10$\mbox{dBm} to show the scaling of performance  considering  light and moderate fog with different pointing errors parameters ($\frac{w_z}{a_r}=\{3, 6\}$ and $\sigma_s=\{5-20\}~{\rm{cm}}$). The figure shows that the OP improves as we increase the number of hops. The number of hops required is less in the light foggy condition than the moderate fog to achieve an acceptable OP of $10^{-3}$.
  
In Fig.~\ref{out_prob_fp}(c), we illustrate the ABER performance for short-range OWC system over the foggy channel with pointing errors. We demonstrate the effect of multihop relaying on the ABER performance for   moderate foggy conditions and  pointing errors ($\frac{w_z}{a_r}=6$ and $\sigma_s=5~{\rm{cm}}$) at two link distances ($\{d=0.8, 1\}~{\rm{km}}$). The figure shows that the ABER  is too high for practical practices in a single hop transmission for moderate fog conditions. However, the use of two relays (i.e., $N=3$ hops) significantly improves the ABER performance of the OWC system.

\section{Conclusion}
We analyzed the performance of a fixed-gain AF relaying based multihop OWC system over atmospheric turbulence and non-zero boresight pointing errors with fog-induced fading. We also analyzed a particular multihop system for short-range communication without considering atmospheric turbulence.  We presented the OP and ABER performance for both multihop systems by deriving closed-form expressions for the PDF and CDF of the end-to-end SNR. We also developed performance metrics for the single-hop system to compare with the multihop transmission.  The diversity order of the considered systems is presented using asymptotic analysis in high SNR to provide better insight into the  performance dependency with various channel and system parameters. We presented extensive simulation results to demonstrate the significance of multihop relaying to extend the communication range. The single-hop transmission does not provide acceptable performance for a longer link over $1500$ \mbox{m} under foggy conditions.  Further, there exists  a  gap between the derived upper bounds and simulation results when the  number of hops increases. However, the OWC system requires a few hops only to achieve acceptable performance for a typical terrestrial communication advocating the significance of analytical bounds. A dual-hop transmission is sufficient to achieve an acceptable outage and error performance in light fog, where a few additional hops  are required to achieve the same performance in moderate fog conditions. 

We envision that the proposed multihop transmission considering generalized fading scenarios would be helpful to assess the deployment of the OWC system for terrestrial  backhaul/fronthaul applications.

\section*{Appendix A: Lemma \ref{pdf_tpf1} (long-range and Single-Hop)}
Applying the theory of product distribution, the PDF of $h_{{tp}_i}=h_{p_i}h_{t_i}$ can be expressed as
\begin{flalign}
f_{h_{{tp}_i}}(x)=\int_{0}^{A_i}\frac{1}{|h_{p_i}|}f_{h_{t_i}}(x/h_{p_i})f_{h_{p_i}}(h_{p_i})dh_{p_i}
\label{pdf_tp_f_dist}
\end{flalign}

We use the identity $\frac{1}{(pt+1)^{q}}=G_{1,1}^{1,1} \left[\begin{matrix} 1-q \\ 0\end{matrix} \bigg| pt \right]$ [\cite{Prudnikov1998}, Eq. (8.4.2.5)] in \eqref{eq:pdf_ht}, we express the PDF of $\cal{F}$ distribution $f_{h_{t_i}}(x)$ by
\begin{flalign}
f_{h_{t_i}}(x)=\frac{a_i^{a_i}x^{a_i-1}}{\beta(a_i,b_i)(b_i-1)^{a_i}}G_{1,1}^{1,1} \left[\begin{matrix} 1-(a_i+b_i) \\ 0\end{matrix} \bigg| Bx \right]
\label{pdf_f_dist_MG}
\end{flalign}

Substituting \eqref{eq:pdf_hp_series} and \eqref{pdf_f_dist_MG} in \eqref{pdf_tp_f_dist}, and applying the definition of Meijer's  G-function, we get 	
\begin{flalign}
	&f_{h_{{tp}_i}}(x)=\frac{a_i^{a_i} \rho_i^2\exp\left(\frac{-s_i^2}{2\sigma_i^2}\right)x^{a_i-1}}{\beta(a_i,b_i)(b_i-1)^{a_i}A_{i}^{\rho_i^2}} \sum_{j=0}^{\infty}\frac{1}{(j!)^2}\left(\frac{s_i^2w^2_{z_{\rm{eq_i}}}}{8\sigma_i^4}\right)^{j}\nonumber\\&\Big(\frac{1}{2\pi \jmath}\int_{\cal{L}}^{}\Gamma(-u)\Gamma((a_i+b_i)+u)\left(B_ix\right)^{u}du\Big) I_1
	\label{pdf_tp_f_dist2}
\end{flalign}
where $I_1=\int_{0}^{A_i}h^{\rho_i^2-a_i-u_i-1}_{p_i} \left(\ln \frac{A_i}{h_{p_i}}\right)^{j} dh_{p_i}$. Substituting $\log \frac{A_i}{h_{p_i}}=t$, we get $I_1=A^{\rho_i^2-a_i-u_i}_i\int_{0}^{\infty}e^{-(\rho_i^2-a_i-u_i)t} t^{j} dt=A^{\rho_i^2-a_i-u_i}_i\frac{\Gamma(1+j)}{(\rho_i^2-a_i-u_i)^{1+j}}=A^{\rho_i^2-a_i-u_i}_i\frac{\Gamma(1+j)[\Gamma(\rho_i^2-a_i-u_i)]^{1+j}}{[\Gamma(1+\rho_i^2-a_i-u_i)]^{1+j}}$. Thus, using $I_1$ in \eqref{pdf_tp_f_dist2} and applying the definition of Meijer G-function, we get the combine PDF of $\cal{F}$-turbulence with the non-zero boresight pointing error
\begin{flalign}
	&f_{h_{{tp}_i}}(x)=\frac{a_i^{a_i} \rho_i^2\exp\left(\frac{-s_i^2}{2\sigma_i^2}\right)}{A^{a_i}_i\beta(a_i,b_i)(b_i-1)^{a_i}} \sum_{j=0}^{\infty}\frac{1}{j!}\left(\frac{s_i^2w^2_{z_{\rm{eq_i}}}}{8\sigma_i^4}\right)^{j}\nonumber\\&x^{a_i-1}G_{2+j,2+j}^{2+j,1} \left[\begin{array}{c} 1-a_i-b_i, \{1+\rho_i^2-a_i\}_0^{j+1}\\ 0, \{\rho_i^2-a_i\}_0^{j+1}\end{array} \left| \frac{Bx}{A_i}\right.\right]
	\label{pdf_tp_f_dist4}
\end{flalign}
Note that 	\eqref{pdf_tp_f_dist4} generalizes the result presented in \cite{Badarneh2021_F} for zero boresight model.

Similarly, we apply the product distribution of two random variables in $h_i=h_{{tp}_i}h_{f_i}$ to get the PDF of the combined channel $f_{h_i}(x)$. Using the transformation on $\gamma_i= \bar{\gamma}_i |h_i|^2$, we  get the PDF of SNR for the combined channel in \eqref{combine_pdf_fpt_series}. We use \eqref{combine_pdf_fpt_series} in $F^{\rm{LR}}_{\gamma_i}(\gamma)=\int_{0}^{\gamma}f^{\rm{LR}}_{\gamma_i}(\gamma)d\gamma$ to derive the CDF of SNR for the combined channel given in \eqref{cdf_fpt_series},  which concludes the proof of Lemma \ref{pdf_tpf1}.
\section*{Appendix B: Theorem \ref{pdf_fpt_series_N} (long-range and Multihop)}
First, we use \eqref{combine_pdf_fpt_series} in \eqref{u-th_moment_snr} and substitute $\sqrt{\gamma}=t$ to get the $u$-th moment of SNR as:
\begin{flalign}
	&\mathbb{E}[\gamma^u_i]=\frac{a_i^{a_i} z_i^{k_i} \rho_i^2\exp\left(\frac{-s_i^2}{2\sigma_i^2}\right)}{(A_i\sqrt{\bar{\gamma}_i})^{a_i}\beta(a_i,b_i)(b_i-1)^{a_i}} \sum_{j=0}^{\infty}\frac{1}{j!}\left(\frac{s_i^2w^2_{z_{\rm{eq_i}}}}{8\sigma_i^4}\right)^{j}\nonumber\\&\int_{0}^{\infty}t^{2(u\frac{\phi_i}{N}+a_i)-1}G_{2+j+k_i,2+j+k_i}^{2+j+k_i,1}\nonumber\\&\Bigg[\begin{array}{c} 1-a_i-b_i, \{1+\rho_i^2-a_i\}_0^{j+1}, \{1+z_i-a_i\}_1^{k_i}\\ 0, \{\rho_i^2-a_i\}_0^{j+1}, \{z_i-a_i\}_1^{k_i}\end{array}\nonumber\\& \left| \frac{B_i}{A_i\sqrt{\bar{\gamma}_i}}t\right.\Bigg]dt
	\label{u-th_moment_snr2}
\end{flalign}

Using the identity \cite[eq. 07.34.21.0009.01]{Wolfram} in \eqref{u-th_moment_snr2}, we get 
\begin{flalign}
	&\mathbb{E}[\gamma^u_i]=\frac{a_i^{a_i} z_i^{k_i} \rho_i^2\exp\left(\frac{-s_i^2}{2\sigma_i^2}\right)}{(A_i\sqrt{\bar{\gamma}_i})^{a_i}\beta(a_i,b_i)(b_i-1)^{a_i}} \sum_{j=0}^{\infty}\frac{1}{j!}\left(\frac{s_i^2w^2_{z_{\rm{eq_i}}}}{8\sigma_i^4}\right)^{j} \nonumber\\&\frac{\big[\Gamma(a_i+\rho^2_i+2\frac{\phi_i}{N}u)\big]^{j+1}\big[\Gamma(a_i+z_i+2\frac{\phi_i}{N}u)\big]^{k_i}\Gamma(2a_i+2\frac{\phi_i}{N}u)}{\left[\Gamma\left(1+a_i+\rho^2_i+2\frac{\phi_i}{N}u\right)\right]^{j+1}\left[\Gamma\left(1+a_i+z_i+2\frac{\phi_i}{N}u\right)\right]^{k_i}}\nonumber\\&\left(\frac{B_i}{A_i\sqrt{\bar{\gamma}_i}}\right)^{-2(u\frac{\phi_i}{N}+a_i)}
	\label{u-th_moment_snr3}
\end{flalign}

Substituting \eqref{u-th_moment_snr3} in \eqref{n_hop_pdf}, applying the definition of Fox's H-function and using the transformation on $\gamma_i= \bar{\gamma}_i |h_i|^2$, we get the PDF of SNR for long-range multihop transmission in \eqref{n_hop_pdf_final}. Integrating the derived PDF $F^{\rm{LR}}_{\gamma_i}(\gamma)=\int_{0}^{\gamma}f^{\rm{LR}}_{\gamma_i}(\gamma)d\gamma$ with the standard mathematical procedure, we get the CDF of end-to-end SNR in \eqref{n_hop_cdf_final}.

\section*{Appendix C: Lemma \ref{pdf_fp_series_shop} (short-range and Single-Hop)}
Using the product distribution in $h_{{fp}_i}=h_{f_i}h_{p_i}$ for the joint PDF of the foggy channel and generalized pointing errors, we represent	
\begin{flalign}
f_{h_{{fp}_i}}(x)=\int_{h_{{fp}_i}/A_i}^{1}\frac{1}{|h_{f_i}|}f_{h_{p_i}}(x/h_{f_i})f_{h_{f_i}}(h_{f_i})dh_{f_i}
\label{combine_pdf_series}
\end{flalign}
Substituting \eqref{eq:pdf_hf} and \eqref{eq:pdf_hp_series} in \eqref{combine_pdf_series}, we get
\begin{flalign}
&f_{h_{{fp}_i}}(x)=\frac{z_i^{k_i}\rho_i^2\exp\left(\frac{-s_i^2}{2\sigma_i^2}\right)}{\Gamma(k_i)A_{i}^{\rho_i^2}}\sum_{j=0}^{\infty}\frac{1}{(j!)^2}\left(\frac{s_i^2w^2_{z_{\rm{eq_i}}}}{8\sigma_i^4}\right)^{j}x^{\rho_i^2-1}\nonumber\\&\int_{h_{{fp}_i}/A_i}^{1}h^{m-1}_{f_i}\left(\log\frac{1}{h_{f_i}}\right)^{k_i-1}\left(\ln\frac{A_ih_{f_i}}{x}\right)^{j}dh_{f_i}
\label{combine_pdf_series1}
\end{flalign}
where $m_i=z_i-\rho_i^2$.
Substituting $\log (1/h_{f_i})=t$ and using binomial expansion $(1-x)^n=\sum_{j=0}^{n}\left( \begin{array}{c} n \\ j \end{array} \right) (1)^{n-j} (-x)^j$ in \eqref{combine_pdf_series1}, and using the definition incomplete Gamma function, we get
\begin{flalign}
&f_{h_{{fp}_i}}(x)=\frac{z_i^{k_i}\rho_i^2\exp\left(\frac{-s_i^2}{2\sigma_i^2}\right)}{\Gamma(k_i)A_{i}^{\rho_i^2}}\sum_{j=0}^{\infty}\frac{1}{(j!)^2}\left(\frac{s_i^2w^2_{z_{\rm{eq_i}}}}{8\sigma_i^4}\right)^{j}\nonumber\\& x^{\rho_i^2-1}\sum_{n=0}^{j}\left( \begin{array}{c} j \\ n \end{array} \right)(-1)^n\left(\ln\frac{A_i}{x}\right)^{j-n}{m_i}^{-n-k_i}\nonumber\\&\left[\Gamma(n+k_i)-\Gamma\left(n+k_i,m_i\ln\frac{A_i}{x}\right)\right]
\label{combine_pdf_series5}
\end{flalign}
Using the transformation $\gamma_i=|h_{{fp}_i}|^2\bar{\gamma}_i$ in \eqref{combine_pdf_series5}, we get the PDF of SNR for OWC system over the foggy channel with generalized pointing errors  in \eqref{eq:combine_pdf_fp_series}. The CDF of SNR for foggy channel with generalized pointing errors can be obtained using \eqref{eq:combine_pdf_fp_series} in $F^{\rm{SR}}_{\gamma_i}(\gamma)=\int_{0}^{\gamma}f^{\rm{SR}}_{\gamma_i}(\gamma)d\gamma$, which results into \eqref{eq:combine_cdf_fp_series}.

\section*{Appendix D:  Theorem \ref{pdf_fp_series_Nhop} (short-range and Multihop)}
Using \eqref{eq:combine_pdf_fp_series} in \eqref{u-th_moment_snr} and substituting $m_i\ln\frac{A_i\sqrt{\bar{\gamma}_i}}{\sqrt{\gamma}}=t$ and  $d\gamma=-\frac{2}{m_i}A^2_i\bar{\gamma}_i\exp\left(-\frac{2}{m_i}t\right)dt$ and applying  the series expansion of incomplete Gamma function $\Gamma(a,t) \triangleq (a-1)! e^{-t} \sum_{l=0}^{a-1} \frac{t^l}{l!}$\cite{Zwillinger2014}, we get 
\begin{flalign}
&\mathbb{E}[\gamma^u_i]=\frac{z_i^{k_i}\rho_i^2\exp\left(\frac{-s_i^2}{2\sigma_i^2}\right)}{m_i^{j+k_i+1}\Gamma(k_i)}(A^2_i\bar{\gamma}_i)^{u\frac{\phi_i}{N}}\sum_{j=0}^{\infty}\frac{1}{(j!)^2}\left(\frac{s_i^2w^2_{z_{\rm{eq_i}}}}{8\sigma_i^4}\right)^{j}\nonumber\\&\sum_{n=0}^{j}\left( \begin{array}{c} j \\ n \end{array} \right)(-1)^n \Big[\Gamma(n+k_i)\int_{0}^{\infty}t^{j-n}\nonumber\\&\exp(-(2u\frac{\phi_i}{m_iN}+\frac{\rho^2_i}{m_i})t)dt-(n+k-1)!\nonumber\\&\sum_{l=0}^{n+k-1}\frac{1}{l! }\int_{0}^{\infty}t^{j-n+l}\exp(-(2u\frac{\phi_i}{m_iN}+\frac{\rho^2_i}{m_i}+1)t)dt\Big]
\label{combine_pdf_fp_series3}
\end{flalign}
To represent the $u$-th moment in terms of Gamma functions, we solve \eqref{combine_pdf_fp_series3} using the identity $\int_{0}^{\infty}t^a \exp(-bt) dt=b^{-a-1}\Gamma(a+1)$ and $\frac{1}{u}=\frac{\Gamma(u)}{\Gamma(1+u)}$ as:

\begin{flalign}
&\mathbb{E}[\gamma^u_i]=\frac{z_i^{k_i}\rho_i^2\exp\left(\frac{-s_i^2}{2\sigma_i^2}\right)}{m_i^{j+k_i+1}\Gamma(k_i)}(A^2_i\bar{\gamma}_i)^{u\frac{\phi_i}{N}}\sum_{j=0}^{\infty}\frac{1}{(j!)^2}\left(\frac{s_i^2w^2_{z_{\rm{eq_i}}}}{8\sigma_i^4}\right)^{j}\nonumber\\&\sum_{n=0}^{j}\left( \begin{array}{c} j \\ n \end{array} \right)(-1)^n \Big[\Gamma(n+k_i)\Gamma(1+j-n)\nonumber\\&\left(\frac{\Gamma\left(\frac{\rho_i^2}{m_i}+\frac{2\phi_i}{m_iN}u\right)}{\Gamma\left(1+\frac{\rho_i^2}{m_i}+\frac{2\phi_i}{m_iN}u\right)}\right)^{n-j-1}-(n+k-1)!\nonumber\\&\sum_{l=0}^{n+k-1}\frac{1}{l! }\Gamma(1+j-n+l)\left(\frac{\Gamma\left(1+\frac{\rho_i^2}{m_i}+\frac{2\phi_i}{m_iN}u\right)}{\Gamma\left(2+\frac{\rho_i^2}{m_i}+\frac{2\phi_i}{m_iN}u\right)}\right)^{n-j-l-1}\Big]
\label{combine_pdf_fp_series4}
\end{flalign}
Thus, invoking \eqref{combine_pdf_fp_series4} in \eqref{n_hop_pdf} and applying the definitions of Fox's H-function, we get PDF of SNR for multihop transmission in \eqref{combine_pdf_fp_series_final}. The  CDF is driven by applying standard mathematical steps in  $F^{\rm{SR}}_{\gamma_i}(\gamma)=\int_{0}^{\gamma}f^{\rm{SR}}_{\gamma_i}(\gamma)d\gamma$ to get   \eqref{combine_cdf_fp_series_new}, which concludes the proof of Theorem \ref{pdf_fp_series_Nhop}.

\bibliographystyle{ieeetran}
\bibliography{bibfile_final_ver1}	
 	\end{document}